\documentclass[12pt,headings=big,numbers=noenddot]{scrartcl}
\usepackage{graphicx,epsfig}
\usepackage{amsmath,amssymb}
\usepackage{array}
\usepackage{arydshln}
\usepackage{color}
\usepackage[nosort]{cite}
\usepackage{enumerate}
\usepackage{subfigure}
\usepackage{bbm}
\usepackage{slashed}
\usepackage{multirow}
\usepackage[bookmarks=false]{hyperref}
\addtokomafont{disposition}{\rmfamily\boldmath}

\newcommand{\be}{\begin{equation}}
\newcommand{\ee}{\end{equation}}
\newcommand{\bea}{\begin{eqnarray}}
\newcommand{\eea}{\end{eqnarray}}
\newcommand{\beq}{\begin{equation}}
\newcommand{\eeq}{\end{equation}}
\newcommand{\beqa}{\begin{eqnarray}}
\newcommand{\eeqa}{\end{eqnarray}}
\newcommand{\ba}{\begin{array}}
\newcommand{\ea}{\end{array}}

\definecolor{nicered}{rgb}{0.7,0.1,0.1}
\newcommand{\ELqL}{\boldsymbol{q}_{\boldsymbol{L}}}
\newcommand{\ELuR}{\boldsymbol{u}_{\boldsymbol{R}}}
\newcommand{\ELdR}{\boldsymbol{d}_{\boldsymbol{R}}}
\newcommand{\EHqL}{q_{3L}}
\newcommand{\EHuR}{t_{R}}
\newcommand{\EHdR}{b_{R}}
\newcommand{\ELqLbar}{\boldsymbol{\bar q}_{\boldsymbol{L}}}

\newcommand{\EHqLbar}{\bar q_{3L}}

\newcommand{\EHdRbar}{\bar b_{R}}

\newcommand{\CLquR}{\boldsymbol{Q_R^u}}

\newcommand{\CHquR}{Q_{3R}^u}

\newcommand{\CLuLbar}{\boldsymbol{\bar U_L}}

\newcommand{\CHuLbar}{\bar T_{L}}

\newcommand{\ELeR}{\boldsymbol{e}_{\boldsymbol{R}}}

\newcommand{\EHeR}{\tau_{R}}
\newcommand{\ELlLbar}{\boldsymbol{\bar l}_{\boldsymbol{L}}}

\newcommand{\EHlLbar}{\bar l_{3L}}

\newcommand{\CLlR}{\boldsymbol{L_R}}

\newcommand{\CHlR}{L_{3R}}

\newcommand{\CLeLbar}{\boldsymbol{\bar E_L}}

\newcommand{\CHeLbar}{\bar{\mathcal{T}}_{L}}

\newcommand{\V}{\boldsymbol{V}}
\newcommand{\Ve}{\boldsymbol{V_e}}

\def\I{\mathcal{I}}

\DeclareFontFamily{OT1}{pzc}{}
\DeclareFontShape{OT1}{pzc}{m}{it}{<-> s * [1.10] pzcmi7t}{}
\DeclareMathAlphabet{\mathpzc}{OT1}{pzc}{m}{it}

\title{\Large\bf\boldmath
Flavour physics from an approximate $U(2)^3$ symmetry
}
\date{\small{E-mail: \href{mailto:barbieri@sns.it}{barbieri@sns.it}, \href{mailto:dario.buttazzo@sns.it}{dario.buttazzo@sns.it}, \href{mailto:filippo.sala@sns.it}{filippo.sala@sns.it}, \href{mailto:david.straub@sns.it}{david.straub@sns.it}}}
\author{\normalsize Riccardo Barbieri, Dario Buttazzo, Filippo Sala, and David M. Straub
\\{\em\normalsize Scuola Normale Superiore and INFN, Piazza dei Cavalieri 7, 56126 Pisa, Italy}
}

\begin{document}

\maketitle
\setcounter{tocdepth}{2}

\begin{abstract}
\noindent
The quark sector of the Standard Model exhibits an approximate $U(2)^3$ flavour symmetry. This symmetry, broken in specific directions dictated by minimality, can explain the success of the Cabibbo-Kobayashi-Maskawa picture of flavour mixing and CP violation, confirmed by the data so far, while allowing for observable deviations from it, as expected in most models of ElectroWeak Symmetry Breaking.
Building on previous work in the specific context of supersymmetry, we analyze the expected effects and we quantify the current bounds in a general Effective Field Theory framework. As a further relevant example we then show how the $U(2)^3$ symmetry and its breaking can be implemented in a generic composite Higgs model and we make a first analysis of its peculiar consequences. We also discuss how  some partial extension of $U(2)^3$ to the lepton sector can arise, both in general and in composite Higgs models.
An optimistic though conceivable interpretation of the considerations developed in this paper
gives  reasons to think that new physics searches in the flavour sector may be about to explore an interesting realm of phenomena.
\end{abstract}

\tableofcontents

\section{Introduction and statement of the program}\label{sec:intro}

The experimental progress of the last decade has shown in a rather spectacular way that the Cabibbo-Kobayashi-Maskawa (CKM) picture of flavour mixing and CP violation in the quark sector, as included in the Standard Model (SM), is fundamentally at work. If one interprets this result as due to a  very short-distance origin of flavour physics   with no expected deviation from the CKM picture close to the Fermi scale, this constitutes a generic problem for theories of ElectroWeak Symmetry Breaking (EWSB) beyond the SM and -- something even more worrisome -- it weakens the hope to be able to provide a rational explanation of quark masses and mixings.
This is not, however, a necessity. An alternative view posits that the success of the CKM picture  be due to the existence of a suitable flavour symmetry appropriately broken in some definite direction, thus allowing a scale of new flavour physics phenomena sufficiently near to the Fermi scale to leave room for relatively small but nevertheless observable deviations from the SM in the flavour sector. An example largely studied in the literature is represented by the hypothesis of Minimal Flavour Violation (MFV), based, in the quark sector, on a $U(3)^3$ symmetry \cite{Chivukula:1987py,Hall:1990ac,D'Ambrosio:2002ex}. Along these lines, here we take the point of view that a more interesting symmetry to  study be a $U(2)_q\times U(2)_u\times U(2)_d$ symmetry -- for brevity $U(2)^3$ -- acting on the first two generations of quarks of different $SU(2)\times U(1)$ quantum numbers, $q_L, u_R$ and $d_R$. The superiority of $U(2)^3$ stems from the observed pattern of quark masses and mixings, which makes  $U(2)^3$ a good approximate symmetry of the SM Lagrangian, broken at most by an amount of order a few $\times 10^{-2}$. This is the size of $V_{cb}$,  comparable to or bigger than the mass ratios $m_{c, u}/m_t$ or $m_{s, d}/m_b$. $U(3)^3$ on the contrary is badly broken at least by the top Yukawa coupling. 

To describe the breaking of $U(2)^3$ we assume that it is encoded in a few small dimensionless parameters. Their origin is unknown and may be different, for example, in different models of EWSB, but we require that they have  definite transformation properties under  $U(2)^3$ itself, so that the overall Lagrangian, fundamental or effective as it may be, remains formally invariant. This is what we mean by saying that $U(2)^3$ is broken in specific directions. Along these lines, the simplest way to give masses to both the up and down quarks of the first two generations is to introduce two (sets of) parameters $\Delta Y_u$, $\Delta Y_d$,  transforming as 
$\Delta Y_u = (2, \bar{2}, 1)$,
$\Delta Y_d = (2, 1, \bar{2})$
under $U(2)_q\times U(2)_u\times U(2)_d$. If these {\it bi-doublets} were the only breaking parameters, the third generation, made of singlets under $U(2)^3$, would not   communicate with the first two generations at all. For this to happen one needs single doublets, at least one,  under any  of the three $U(2)$'s. The only such doublet that can explain the observed small mixing between the third and the first two generations, in terms of a correspondingly small parameter, transforms under $U(2)_q\times U(2)_u\times U(2)_d$ as 
$ \V = (2,1,1)$.
A single doublet under $U(2)_u$ or $U(2)_d$ instead of $U(2)_q$ would have to be of order unity. 
To summarize, we assume that $U(2)^3$ is an approximate symmetry of the flavour sector of the SM only weakly broken in the directions (by the {\it spurions})
\begin{equation}
\Delta Y_u = (2, \bar{2}, 1),~~
\Delta Y_d = (2, 1, \bar{2}),~~\V = (2,1,1).
\label{directions}
\end{equation}

The consequences of this framework for supersymmetric extensions of the SM has already been examined in Ref.~\cite{Barbieri:2011ci,Barbieri:2011fc}\footnote{A $U(2)^3$ symmetry with a {\em non-minimal} breaking pattern has recently been considered in the context of radiative radiative flavour violation in the MSSM \cite{Crivellin:2011sj}.}. Within the same general framework, in this work we aim to address the following questions:
\begin{itemize}
\item Taking an Effective Field Theory point of view, which are the limits on the possible size of new flavour changing and CP violating interactions consistent with the current observations? 
 Which signals can one in turn expect in foreseen experiments?

\item How can one implement the $U(2)^3$ picture in a theory of EWSB containing a composite Higgs boson? 

\item Can there be at least some partial extension of $U(2)^3$ to the lepton sector, notwithstanding the peculiar mixing properties shown by neutrinos, apparently quite different from the ones of the quarks?
\end{itemize}
While addressing each of these questions in sections \ref{sec:eft}, \ref{sec:composite} and \ref{sec:leptons}, respectively, it will be interesting to compare the answers with those one would obtain from a $U(3)^3$ symmetry.

\section{$U(2)^3$ in Effective Field Theory}\label{sec:eft}

\subsection{Relevant effective operators in the physical quark basis}

Given the basic distinction between the third and the first two generations of quarks, we adopt for the triplets of left-handed  doublets and right-handed charge $2/3$ and $-1/3$ quark singlets respectively the self-explanatory notation
\begin{equation}
q_L = \begin{pmatrix} \ELqL \\ \EHqL\end{pmatrix},\qquad
u_R = \begin{pmatrix} \ELuR \\ \EHuR\end{pmatrix},\qquad
d_R = \begin{pmatrix} \ELdR \\ \EHdR\end{pmatrix},
\label{BoldNotation}
\end{equation}
All the relevant operators are built from chirality conserving or chirality breaking quark bilinears:
\begin{equation}
\bar{q}_{Li}\gamma_\mu X_{ij}q_{Lj} \approx
a\, \EHqLbar \gamma_\mu \EHqL +
b\, \ELqLbar\gamma_\mu \ELqL +
c\, \EHqLbar\gamma_\mu (\V^\dagger \ELqL) +
c^*\, (\ELqLbar \V) \gamma_\mu \EHqL +
d \,(\ELqLbar \V) \gamma_\mu (\V^\dagger \ELqL),
\label{cc}
\end{equation}
\begin{equation}
\bar{q}_{Li}X_{ij}^{(d)}d_{Rj} \approx \lambda_b \left(a_d\, \EHqLbar  \EHdR + b_d\, (\ELqLbar \V) \EHdR + c_d\, \ELqLbar \Delta Y_d \ELdR\right),
\label{cb_d}
\end{equation}
\begin{equation}
\bar{q}_{Li}X_{ij}^{(u)}u_{Rj} \approx \lambda_t \left(a_u\, \EHqLbar  \EHuR + b_u\, (\ELqLbar \V) \EHuR + c_u\, \ELqLbar \Delta Y_u \ELuR\right).
\label{cb_u}
\end{equation}
In each of these equations we have neglected higher order terms in $\V, \Delta Y_u, \Delta Y_d$ which, unless differently specified, do not affect the following considerations and we have included order 1 coefficients, $a, b, c, \dots$, dependent upon the specific bilinear under consideration.  Except for $a, b, d$ in Eq. (\ref{cc}), which are constrained to be real by hermiticity, all these coefficients are in general complex.  In each  of the chirality breaking bilinears, relevant both for the mass terms and for the $\sigma_{\mu \nu}$-terms,  we have factored out the parameters $\lambda_t \approx m_t/v$ and $\lambda_b\approx m_b/v$. While $\lambda_t$ is of order unity, the smallness of $\lambda_b$ may be attributed to an approximate $U(1)_d$ acting on all the right-handed down quarks in the same way inside and outside $U(2)^3$. In presence of more than one Higgs doublet, $\lambda_b$ can get bigger values. Finally there is no significant  flavour violation in the chirality conserving right-handed quark bilinears.

 It is useful to reduce by sole $U(2)^3$ transformations the parameters $\V, \Delta Y_u, \Delta Y_d$ to the form
\begin{equation}\label{spurions}
\V^T = (0,~ \epsilon),~~\Delta Y_u = R_{12}^u \,\Delta Y_u^\text{diag},~~\Delta Y_d = U_{12}^d \,\Delta Y_d^\text{diag},
\end{equation}
where $\epsilon$ is a real parameter of the same size as $|V_{cb}|$, $R_{12}$ is a rotation matrix in the space of the first two generations, $U_{12}^d$ is a unitary matrix of the form $U_{12}^d = \Phi_1 R_{12}^d$ with $\Phi_1 = \text{diag}(e^{i\phi}, 1)$, and $\Delta Y_u^\text{diag}$, $\Delta Y_d^\text{diag}$ are real diagonal matrices. Incidentally this shows that, if CP violation only resides in $\V, \Delta Y_u, \Delta Y_d$, there is a single physical phase, $\phi$, which gives rise to the CKM phase. To obtain the relevant flavour changing and CP violating effective operators in the physical quark basis, one has to reduce the kinetic terms to canonical form and the mass matrices for the up and down quarks to real diagonal form.  Due to the specific breaking of $U(2)^3$ in the directions~(\ref{directions}), a crucial fact is that the mass matrices are diagonalized, to a sufficient level of accuracy, by transformations acting on the left-handed fields only.

As a result one gets (see Appendix~\ref{sec:yukawa} for details)\footnote{In the following all the matrices are in three-dimensional space, with $R_{12}^u$ and $U_{12}^d$ extended with a $1$ in the $3 3$ position.}:
\begin{enumerate}[i)]
\item for the CKM matrix:
\begin{equation}\label{CKM}
\bar{u}_{Li}\gamma_\mu V_{ij}d_{Lj},~~ V \approx (R_{12}^u)^T U_{23}^\epsilon U_{12}^d,
\end{equation}
where $U_{23}^\epsilon$ is a unitary matrix in the $2 3$ sector with off-diagonal elements of order $\epsilon$.

\item for the chirality breaking bilinears in the down sector:
\begin{equation}\label{cbdown}
\bar{d}_{Li}\sigma_{\mu \nu} \mu_{ij}^\alpha d_{Rj},~~ 
\mu^\alpha  \approx \lambda_b \left(A^\alpha (U_{12}^d)^\dagger  U_{23}^\alpha \mathcal{I}_3 + B^\alpha \Delta \widetilde Y_d^\text{diag}\right),
\end{equation}
where $A^\alpha, B^\alpha$ are parameters of order unity, in general complex, $\mathcal{I}_3 = \text{diag}(0, 0, 1)$ and $\Delta \widetilde Y_d^\text{diag} = \text{diag}(\epsilon_1^d, \epsilon_2^d, 0)$, with $\epsilon_1^d$ and $\epsilon_2^d$ the diagonal entries of $\Delta Y_d^\text{diag}$.

\item for the chirality conserving bilinears:
\begin{equation}
\bar{d}_{Li}\gamma_{\mu} X_{ij}^\beta d_{Lj},~~ 
X^\beta \approx  A^\beta \mathbf{1} + B^\beta (U_{12}^d)^\dagger  U_{23}^\beta \mathcal{I}_{3 2} (U_{23}^\beta)^\dagger  U_{12}^d,
\end{equation}
where  $A^\beta, B^\beta$ are real parameters and  $\mathcal{I}_{3 2} = {\rm diag}(0, O(\epsilon^2), 1)$. $U_{23}^{\alpha,\beta}$ are unitary matrices in the $2 3$ sector with off-diagonal elements of order $\epsilon$ and, as $U_{23}^\epsilon$, are real if CP violation only resides in the spurions.
\end{enumerate}

Finally this leads to the following exhaustive set of relevant flavour changing effective operators\footnote{We omit $\Delta F=1$ operators which only enter low-energy observables in a fixed linear combination with the ones considered here and do not lead to qualitatively new effects.}, all weighted  by the square of an inverse mass scale $1/\Lambda$:

\begin{enumerate}[i)]
\item $\Delta B =2, i = s,d$
\begin{equation}
c_{LL}^B e^{i \phi_B} \xi_{ib}^2 \frac{1}{2}(\bar{d}_{Li}\gamma_\mu b_L)^2,
\label{eff1}
\end{equation}

\item $\Delta S =2$
\begin{equation}
c_{LL}^K  \xi_{ds}^2 \frac{1}{2}(\bar{d}_{L}\gamma_\mu s_L)^2,
\end{equation}

\item $\Delta B =1, i = s,d$, chirality breaking ($\alpha = \gamma, G$):
\begin{equation}
c^\alpha e^{i \phi^\alpha} \xi_{ib} m_b (\bar{d}_{Li}\sigma_{\mu \nu} b_R)  O^\alpha_{\mu \nu}, ~~
O^\alpha_{\mu \nu}= e F_{\mu\nu}, ~ g_s G_{\mu \nu},
\end{equation}

\item $\Delta B =1, i = s,d$, chirality conserving ($\beta = L, R, H$):
\begin{equation}
c^\beta_B e^{i\phi^ \beta} \xi_{ib} (\bar{d}_{Li}\gamma_\mu b_L) O^\beta_\mu, ~~
O^\beta_\mu = (\bar{l}_L\gamma_\mu l_L),~(\bar{e}_R\gamma_\mu e_R),~(H^\dagger D_\mu H),
\end{equation}

\item $\Delta S =1$, chirality conserving:
\begin{equation}
c^\beta_K  \xi_{ds}(\bar{d}_{L}\gamma_\mu s_L)O^\beta_\mu, ~~
O^\beta_\mu = (\bar{l}_L\gamma_\mu l_L),~(\bar{e}_R\gamma_\mu e_R),~(H^\dagger D_\mu H),
\label{eff5}
\end{equation}
\end{enumerate}
where $\xi_{ib}= V_{tb}V_{ti}^*$, $\xi_{ds}= V_{ts}V_{td}^*$ and  the coefficients 
$c_{LL}^B, c_{LL}^K, c^\alpha, c^\beta_B, c^\beta_K$ are  real, with the phases made explicit wherever present. All these coefficients are model dependent and, in principle, of similar order.

Analogous expressions can be found for operators involving the up quarks. However if these operators are weighted by the same scale $\Lambda$ as for the down quarks, they are phenomenologically irrelevant unless some of the relative dimensionless coefficients are at least one order of magnitude bigger than the ones in Eq.~(\ref{eff1}-\ref{eff5}). This is in particular the case for operators contributing to $D-\bar{D}$ mixing, to direct CP violation in $D$-decays or to top decays, $t\rightarrow c \gamma$ or $t\rightarrow c Z$ (see Appendix~\ref{sec:up}).

\subsection{Comparison with $U(3)^3$ at small and large $\tan{\beta}$}

It is useful to compare the expectations of the $U(2)^3$ symmetry suitably broken as described in Section~\ref{sec:intro} with the case of $U(3)_Q\times U(3)_u\times U(3)_d$ broken in the directions $Y_u= (3, \bar{3},1)$ and $Y_d = (3, 1, \bar{3})$, as we now briefly recall from the literature \cite{D'Ambrosio:2002ex,Feldmann:2008ja,Kagan:2009bn,Isidori:2012ts}.

First, by sole $U(3)^3$ transformations, one can set without loss of generality
\begin {equation}
Y_u = V_0^\dagger  Y_u^\text{diag}, ~~Y_d = Y_d^\text{diag},
\label{yud_3}
\end{equation}
where $ Y_u^\text{diag}, Y_d^\text{diag}$ are real diagonal matrices and $V_0$ is a unitary matrix\footnote{We use the notation $V_0$ to distinguish it from the CKM matrix $V$ which differs in general by order one corrections from $V_0$.} dependent on one single phase. Therefore the CKM phase is the only source of CP violation if no new phase is born outside of $Y_u$ or $Y_d$, no matter what the value of $\tan{\beta}$ is. 

As in the $U(2)^3$ case, to determine the relevant flavour violating operators one has to reduce the kinetic terms to canonical form and the mass matrices to real diagonal form. In turn this depends on the value of $\tan{\beta}$ which determines the need to include or not powers of $Y_d Y_d^\dagger $  in effective operators. For moderate $\tan{\beta}$ the relevant flavour violating quark bilinears in the physical basis have the approximate form ($ i,j = d,s,b$):
\begin{equation}
A \,\bar{d}_{Li}\sigma_{\mu \nu}(V^\dagger  \mathcal{I}_3 V Y_d^\text{diag})_{ij}d_{Rj},~~
B \,\bar{d}_{Li}\gamma_{\mu}  (V^\dagger  \mathcal{I}_3 V)_{ij} d_{Lj},
\end{equation}
where $A$ is a complex parameter and $B$ a real one. This leads to the following set of relevant operators ($\xi_{ij} = V_{tj} V_{ti}^*$):

i) $\Delta F =2$
\begin{equation}
C_{LL}  \xi_{ij}^2 \frac{1}{2}(\bar{d}_{Li}\gamma_\mu d_{Lj})^2,
\end{equation}

ii) $\Delta F =1$, chirality breaking ($\alpha = \gamma, G$):
\begin{equation}
C^\alpha e^{i \chi^\alpha} \xi_{ij}  (\bar{d}_{Li}\sigma_{\mu \nu} m_j d_{Rj})  O^\alpha_{\mu \nu}, ~~
O^\alpha_{\mu \nu}= e F_{\mu\nu}, ~ g_s G_{\mu \nu},
\end{equation}

iii) $\Delta F =1$, chirality conserving ($\beta = L, R, H$):
\begin{equation}
C^\beta  \xi_{ij}(\bar{d}_{Li}\gamma_\mu d_{Lj})O^\beta_\mu, ~~
O^\beta_\mu = (\bar{l}_L\gamma_\mu l_L),~(\bar{e}_R\gamma_\mu e_R),~(H^\dagger D_\mu H),
\end{equation}
with $C_{LL}, C^\alpha, C^\beta$ real.

At large $\tan{\beta}$  both powers of $Y_u Y_u^\dagger $ and $Y_d Y_d^\dagger $ are relevant in effective operators \cite{Feldmann:2008ja,Kagan:2009bn,Colangelo:2008qp}. The fact that  $\lambda_t$ and $\lambda_b$ are both of $O(1)$ leads to the the breaking of $U(3)^3$ down to $U(2)^3$. Note  that this is not enough to conclude that $U(3)^3$ at large  $\lambda_b$ leads to the same pattern of flavour violation described in the previous subsection. For this to happen one needs that the breaking directions of $U(2)^3$ be as in (\ref{directions}). However, due to the fact that $U(3)^3$ is only broken in the $Y_{u, d}$ directions, this is  the case. To see this, after suitable $U(3)^3$ transformations, one can write $Y_{u, d}$ as \cite{Feldmann:2008ja,Kagan:2009bn}
\begin{equation}
Y_{u,d} =e^{\pm i \hat{\chi}/2}
\,\lambda_{t,b}
\begin{pmatrix}
\Delta Y_{u,d} & 0 \\
0 & 1
\end{pmatrix},
\end{equation}
where  $\Delta Y_{u,d}$ are $2\times 2$ matrices as in (\ref{directions}) and $ \hat{\chi}$ is a  hermitian $3\times 3$ matrix with $\chi$ a 2-component vector,
\begin{equation}
 \hat{\chi} =
\begin{pmatrix}
0 & \boldsymbol{\chi} \\
\boldsymbol{\chi}^\dagger & 0
\end{pmatrix},
\end{equation}
which determines the misalignement of $Y_{u,d}$  in the $1 3$ and $2 3$ directions, known to be small, of order $|V_{cb}|$, from the CKM matrix. Expanding in $ \hat{\chi}$, 
\begin{equation}
Y_{u,d} \approx
\lambda_{t,b}
\begin{pmatrix}
\Delta Y_{u,d} & \pm i\boldsymbol{\chi}/2 \\
\pm i\boldsymbol{\chi}^\dagger\Delta Y_{u,d}/2 & 1
\end{pmatrix},
\label{yudmatrix}
\end{equation}
which shows that $\boldsymbol{\chi}$ plays the same role as $\V$. For example the term in the bottom left of (\ref{yudmatrix}), of subleading order, is given in the language of (\ref{cb_d}, \ref{cb_u}) by $\bar{q}_3 (\V^\dagger  \Delta Y_d \ELdR)$ or $\bar{q}_3 (\V^\dagger  \Delta Y_u \ELuR)$\footnote{As in the case of Eq.~(\ref{yud_3}), note that $Y_{u,d}$ in (\ref{yudmatrix}) are not in general proportional to the physical $u, d$ mass matrices, due to the presence of $Y_uY_u^\dagger $ and $Y_dY_d^\dagger $ corrections both in the mass terms themselves and in the kinetic terms.}.
Therefore $U(3)^3$ at large $\tan{\beta}$ leads in general to the same effective operators as in (\ref{eff1}-\ref{eff5}), apart from the characteristic $\tan{\beta}$-dependence of the various coefficients, which should also show up both in flavour violating and in suitable flavour conserving amplitudes.

\subsection{Current bounds and possible new effects}

\subsubsection{$\Delta F=2$ processes}

As discussed above, the relevant $\Delta F=2$ operators generated in the $U(2)^3$ framework read
\begin{equation}
\mathcal H_{\rm eff} =
\frac{c_{LL}^K }{\Lambda^2} \xi_{ds}^2 \frac{1}{2}\left({\bar d}_L \gamma_\mu  s_L \right)^2
+\sum_{i=d,s}
\frac{c_{LL}^B e^{i\phi_B}}{\Lambda^2} \xi_{ib}^2\frac{1}{2}\left({\bar d}_L^i \gamma_\mu  b_L \right)^2
+\text{h.c.}\,,
\label{eq:HeffDF2}
\end{equation}
where $c_{LL}^{K,B}$ are real, model dependent parameters that can be of $O(1)$. The $U(3)^3$ case at low $\tan\beta$ is recovered for $c_{LL}^K=c_{LL}^B$ and $\phi_B=0$. As shown in \cite{Barbieri:2011ci}, the observables in $K$, $B_d$ and $B_s$ meson mixing are modified as
\begin{align}
\epsilon_K&=\epsilon_K^\text{SM(tt)}\left(1+h_K\right) +\epsilon_K^\text{SM(tc+cc)} ,
\label{eq:epsKxF}\\
S_{\psi K_S} &=\sin\left(2\beta + \text{arg}\left(1+h_B e^{i\phi_B}\right)\right) ,
\label{eq:Spk} \\
S_{\psi\phi} &=\sin\left(2|\beta_s| - \text{arg}\left(1+h_B e^{i\phi_B}\right)\right) ,
\label{eq:SpsiphixF}\\
\Delta M_d &=\Delta M_d^\text{SM}\,\left|1+h_B e^{i\phi_B}\right| ,
\label{eq:DMdxF}\\
\frac{\Delta M_d}{\Delta M_s} &= \frac{\Delta M_d^\text{SM}}{\Delta M_s^\text{SM}} \,,
\label{eq:MdMs}
\end{align}
where
\begin{align}
h_{K,B} &= c_{LL}^{K,B}
\frac{4s_w^4}{\alpha_{em}^2S_0(x_t)}
\frac{m_W^2}{\Lambda^2}
\approx
1.08\,c_{LL}^{K,B}
\left[ \frac{3\,\text{TeV}}{\Lambda} \right]^2
\,.
\end{align}
In the special case of supersymmetry with dominance of gluino contributions, one has $h_K=x^2F_0>0$ and $h_B=x F_0$, where $F_0$ is a positive loop function given in \cite{Barbieri:2011ci} and $x$ an $O(1)$ mixing parameter.

To confront the effective Hamiltonian (\ref{eq:HeffDF2}) with the data, the dependence of the $\Delta F=2$ observables on the CKM matrix elements has to be taken into account. To this end, we performed global fits of the CKM Wolfenstein parameters $A$, $\lambda$, $\bar\rho$ and $\bar\eta$ as well as the coefficients $c_{LL}^{K,B}$ and the phase $\phi_B$ to the set of experimental observables collected in the left-hand column of Tab.~\ref{tab:inputs}, by means of a Markov Chain Monte Carlo, assuming all errors to be Gaussian.

\begin{figure}[tb]
\centering
\includegraphics[width=\textwidth]{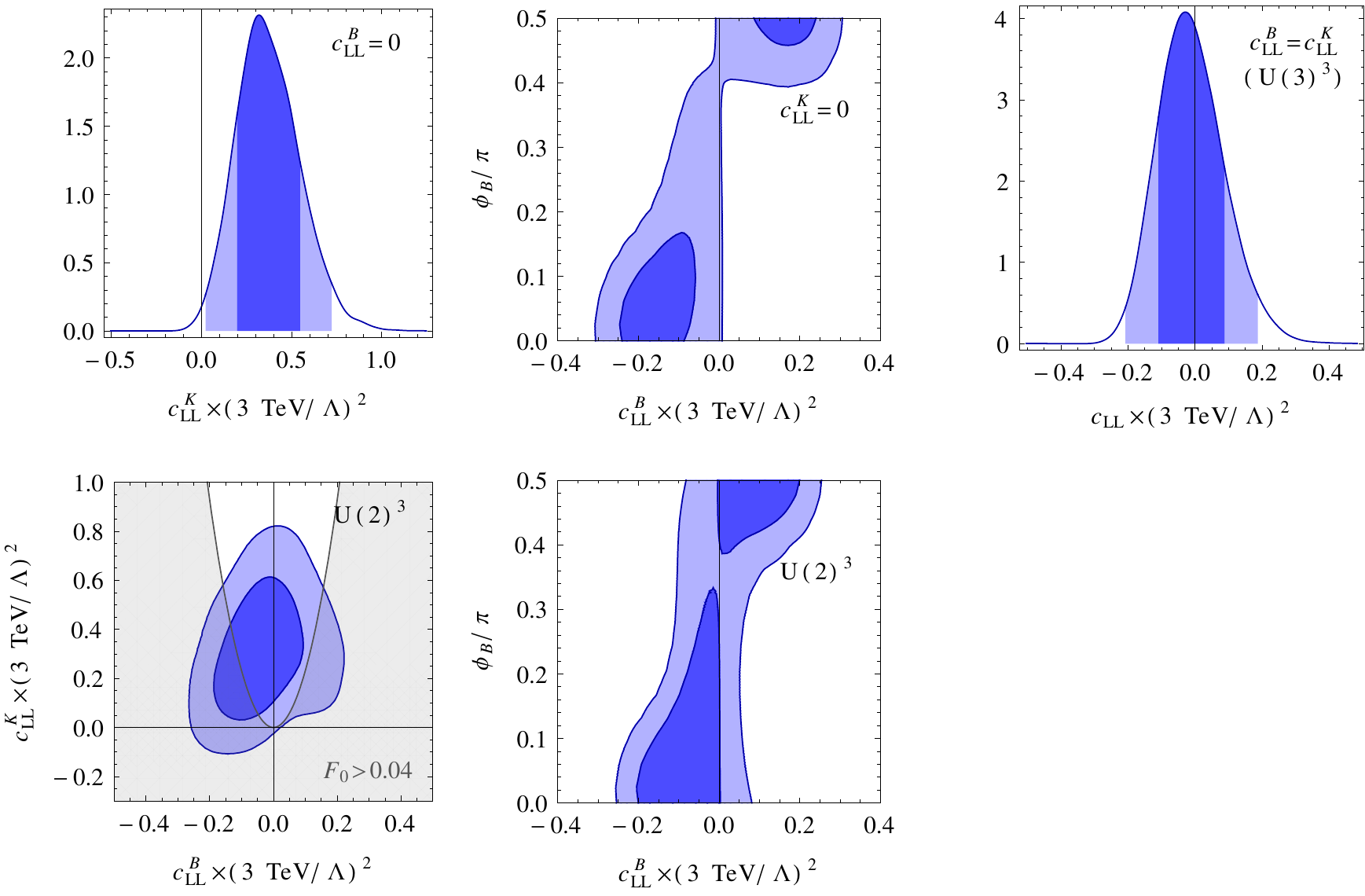}
\caption{Fit predictions (68 and 95\% Bayesian credible regions) in $\Delta F=2$ fits with $c_{LL}^B=0$ (top left), $c_{LL}^K=0$ (top centre), $c_{LL}^B=c_{LL}^K$, $\phi_B=0$ (top right, relevant to $U(3)^3$) and with all 3 parameters independent (bottom). The gray region in the bottom left plot is disfavoured by direct searches only in the SUSY case with dominance of gluino contributions.}
\label{fig:DF2fits}
\end{figure}

The results of four different fits are shown in Fig.~\ref{fig:DF2fits}. The top left panel shows the fit prediction for $c_{LL}^K$ in a fit with $c_{LL}^B=0$. The top centre panel shows the fit prediction in the $c_{LL}^B$--$\phi_B$ plane in a fit with $c_{LL}^K=0$. The preference for non-SM values of the parameters in both cases arises from the tension in the SM CKM fit between $\epsilon_K$ (when using the experimental data for $V_{cb}$ and $\sin2\beta$ as inputs) and $S_{\psi K_S}=\sin2\beta$ \cite{Lunghi:2008aa,Buras:2008nn,Altmannshofer:2009ne,Lunghi:2010gv,Bevan:2010gi,Brod:2011ty}. As is well known, this tension can be solved either by increasing $\epsilon_K$ (as in the first case) or by decreasing $S_{\psi K_S}$ by means of a new physics contribution to the $B_d$ mixing phase (as in the second case). In the second case, also a positive contribution to $S_{\psi\phi}=-\sin2\phi_s$ is generated.
The top right panel shows the fit prediction in a fit where $\phi_B=0$ and $c_{LL}^B=c_{LL}^K\equiv c_{LL}$, i.e. the $U(3)^3$ or MFV limit. In that case, a positive $c_{LL}$ cannot solve the CKM tension, since it would lead to an increase not only in $\epsilon_K$, but also in $\Delta M_{d,s}$.

The two plots in the bottom row of Fig.~\ref{fig:DF2fits} show the projections onto the $c_{LL}^K$--$c_{LL}^B$ and $c_{LL}^B$--$\phi_B$ planes of the fit with all 3 parameters in (\ref{eq:HeffDF2}) non-zero. Since both solutions to the CKM tension now compete with each other, the individual parameters are less constrained individually.
In the case of supersymmetry with dominance of gluino contributions, as mentioned above, the modification of the $B_{d,s}$ and $K$ mixing amplitudes is correlated by the common loop function $F_0$, which depends on the gluino and left-handed sbottom masses. Taking into account direct constraints from LHC on the sbottom and gluino masses, one finds that values of $F_0$ above about 0.04 are disfavoured. This constraint is shown in the bottom-left panel of Fig.~\ref{fig:DF2fits} as a gray region. However, we note that this bound (but not the correlation predicted by $U(2)^3$) is invalidated once chargino contributions dominate.

\begin{table}[tb]
\renewcommand{\arraystretch}{1.0}
 \begin{center}
\begin{tabular}{llllll}
\hline
$|V_{ud}|$ & $0.97425(22)$ &\cite{Hardy:2008gy}& $f_K$  & $(155.8\pm1.7)$ MeV & \cite{Laiho:2009eu}\\
$|V_{us}|$ & $0.2254(13)$ &\cite{Antonelli:2010yf}& $\hat B_K$ & $0.737\pm0.020$ &\cite{Laiho:2009eu} \\
$|V_{cb}|$ & $(40.6\pm1.3)\times10^{-3}$ &\cite{Nakamura:2010zzi}& $\kappa_\epsilon$ & $0.94\pm0.02$ & \cite{Buras:2010pza}\\
$|V_{ub}|$ & $(3.97\pm0.45)\times10^{-3}$ &\cite{Kowalewski:2011zz}& $f_{B_s}\sqrt{\hat B_s}$  & $(288\pm15)$ MeV &\cite{Lunghi:2011xy}\\
$\gamma_{\rm CKM}$ & $(74\pm11)^\circ$ &\cite{Bevan:2010gi}& $\xi$ & $1.237\pm0.032$ &\cite{Laiho:2009eu}\\
$|\epsilon_K|$ & $(2.229\pm0.010)\times10^{-3}$ &\cite{Nakamura:2010zzi} &$\eta_{tt}$&$0.5765(65)$&\cite{Buras:1990fn}\\ 
$S_{\psi K_S}$ & $0.673\pm0.023$ &\cite{Asner:2010qj} &$\eta_{ct}$&$0.496(47)$&\cite{Brod:2010mj}\\
$\Delta M_d$ & $(0.507\pm0.004)\,\text{ps}^{-1}$ &\cite{Asner:2010qj} &$\eta_{cc}$&$1.38(53)$&\cite{Brod:2011ty}\\
$\Delta M_s/\Delta M_d$ & $(35.05\pm0.42)$ &\cite{Abulencia:2006ze,Asner:2010qj} &&&\\
$\phi_s$ & $-0.002 \pm 0.087$ & \cite{LHCb-TALK-2012-029} &&&\\
\hline
 \end{tabular}
 \end{center}
\caption{Observables and hadronic parameters used as input to the $\Delta F=2$ fits.}
\label{tab:inputs}
\end{table}

\subsubsection{$\Delta F=1$ processes}

The $U(2)^3$ predictions for $\Delta F=1$ processes are more model-dependent because a larger number of operators is relevant. In addition, the main prediction of universality of $b\to s$ and $b\to d$ amplitudes --  but not $s\to d$ amplitudes -- is not well tested. Firstly, current data are better for $b\to s$ decays compared to $b\to d$ decays. Secondly, the only clean $s\to d$ processes are $K\to\pi\nu\bar\nu$ decays, but $b\to q\nu\bar\nu$ processes have not been observed yet.
Thus, in the following we will present the constraints on the effective Hamiltonian ($v = 246$ GeV)
\begin{multline}
\mathcal H_{\rm eff} =\sum_{i=d,s} \xi_{ib}
\bigg[
\frac{c_{7\gamma}e^{i\phi_{7\gamma}}}{\Lambda^2}m_{b} \left( {\bar d^i}_L \sigma_{\mu\nu}  b_R\right) e F^{\mu\nu}
+
\frac{c_{8g}e^{i\phi_{8g}}}{\Lambda^2}m_{b} \left( {\bar d^i}_L \sigma_{\mu\nu} T^a b_R \right) g_s G^{\mu\nu\,a}
\\
+
\frac{c_Le^{i\phi_{L}}}{\Lambda^2} \left( {\bar d}_L^i \gamma_{\mu} b_L \right) \left( {\bar l}_L \gamma_{\mu} l_L \right)+
\frac{c_Re^{i\phi_{R}}}{\Lambda^2} \left( {\bar d}_L^i \gamma_{\mu} b_L \right) \left( {\bar e}_R \gamma_{\mu} e_R \right)+
\frac{c_He^{i\phi_{H}}}{\Lambda^2} \frac{v^2}{2} \left( {\bar d}_L^i  \gamma_\mu b_L \right)\frac{g}{c_w}Z^{\mu}
\bigg]
+\text{h.c.}\,,
\label{eq:HeffDF1}
\end{multline}
using only data from inclusive and exclusive $b\to s$ decays, making use of the results of \cite{Altmannshofer:2011gn}.
While in the supersymmetric case, only the coefficients $c_{7\gamma}$ and $c_{8g}$ are relevant \cite{Barbieri:2011fc}, in general all the coefficients in~(\ref{eq:HeffDF1}) can be relevant and of $O(1)$.
Since the chromomagnetic penguin operator enters the observables considered in the following only through operator mixing with the electromagnetic one, we will ignore $c_{8g}$ in the following.

\begin{figure}[tb]
\centering
\includegraphics[width=0.68\textwidth]{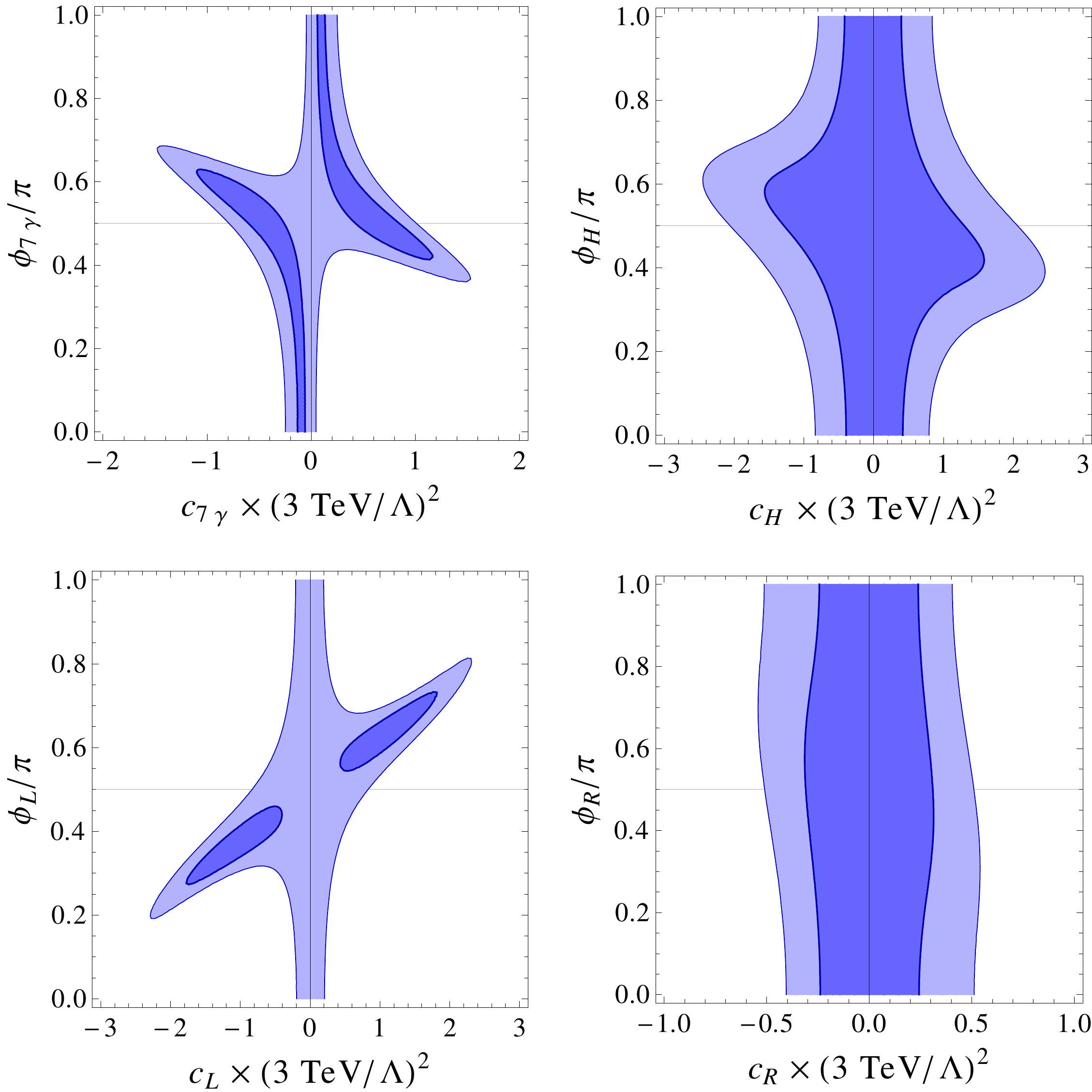}
\caption{68 and 95\% C.L. allowed regions for the $\Delta F=1$ coefficients in $U(2)^3$, using the results of a global analysis of inclusive and exclusive $b\to s$ decays \cite{Altmannshofer:2011gn}.}
\label{fig:DF1fits}
\end{figure}

Fig.~\ref{fig:DF1fits} shows the constraints on the coefficients of the four operators in~(\ref{eq:HeffDF1}) and their phases. The constraints are particularly strong for the magnetic penguin operator and the semi-leptonic left-left vector operator. In the first case, this is due to the $B\to X_s\gamma$ branching ratio, in the second case due to the forward-backward asymmetry in $B\to K^*\mu^+ \mu^-$. Interestingly, in both cases, the constraint is much weaker for maximal phases of the new physics contribution, since the interference with the (real) SM contribution is minimized in that case. For the two operators in the right-hand column of Fig.~\ref{fig:DF1fits}, this effect is less pronounced. The reason is that both coefficients are accidentally small in the SM: in the first case due to the small $Z$ coupling to charged leptons proportional to $(1-4s_w^2)$ and in the second case due to $C_9(\mu_b)\approx-C_{10}(\mu_b)$ (in the convention of  \cite{Altmannshofer:2011gn}).

To ease the interpretation of Fig.~\ref{fig:DF2fits},~\ref{fig:DF1fits}, all the coefficients are normalized, as explicitly indicated, to a scale $\Lambda = 3$ TeV, which might represent both the scale of a new strong interaction responsible for EWSB, $\Lambda_s \approx 4\pi v$, or the effective scale from loops involving the exchange of some new weakly interacting particle(s) of mass of $O(v)$. Interestingly the fits of the current flavour data are generally consistent with coefficients of order unity, at least if there exist sizable non vanishing phases, when they are allowed. Note in this respect that $U(3)^3$ at low $\tan{\beta}$ does not allow phases in $c_{L, R, H}$ in Fig.~\ref{fig:DF1fits}, with correspondingly more stringent constraints especially on $c_L$. A possible interpretation of these Fig.~\ref{fig:DF2fits},~\ref{fig:DF1fits} is that the current flavour data are at the level of probing the $U(2)^3$ hypothesis in a region of parameter space relevant to several new theories of EWSB. Needless to say, the presence of phases in flavour-diagonal chirality breaking operators has to be consistent with the limits coming from the neutron Electric Dipole Moment. It should also be noted that the coefficients of the flavour diagonal operators, analogue of the ones in the second line of \eqref{eq:HeffDF1}, are limited by the ElectroWeak Precision Tests at a similar level to the ones shown in Fig. \ref{fig:DF1fits}.

\section{$U(2)^3$ in composite Higgs models}\label{sec:composite}

\subsection{General setup}

If the Higgs boson is a composite particle by a suitable new strong interaction, the problem of describing flavour in a way consistent with current experiments is non trivial. It is therefore natural to ask if $U(2)^3$ can be implemented in this case as well and  to see which are its general consequences, if the implementation is possible at all. 

At least in an effective description the way the standard fermions (collectively called $f$) see the composite Higgs boson(s), $H$, is by admitting the presence of composite vector-like fermions ($F$), naturally coupled to $H$ by the strong interactions through Yukawa-like terms, schematically  $\lambda FHF$, and mixed with the standard fermions by mass terms, $m Ff$ \cite{Kaplan:1991dc,Contino:2006nn}. Both $Y$ and $m$ are assumed to break in specific directions the global flavour symmetries acting on $F$ and $f$.

What does one need to try to get a consistent description of flavour in this specific setup? One way introduced in \cite{Barbieri:2008zt}\footnote{But actually effectively implemented before in a 5-dimensional example in \cite{Cacciapaglia:2007fw}.} is to assume a sufficiently large flavour symmetry respected by the strong sector, including the coupling $\lambda  FHF$, and shared by some of the elementary fermions. To see explicitly how this works, let us consider the $U(3)^3$ example recently discussed in \cite{Redi:2011zi}.

The vector-like fermions $F$ are composed of two $SU(2)_L$ doublets, $Q^u$ and $Q^d$, with the same quantum numbers under the SM gauge group as the standard $q_L$, and of $SU(2)_L$ singlets, $U$ and $D$, with the quantum numbers of $u$ and $d$ respectively\footnote{Each of the composite fermions may be part of a larger multiplet, so that, e.g., the Yukawa coupling $\lambda  FHF$ enjoys a  $SU(2)_L\times SU(2)_R$ symmetry. This is not of relevance, however, to the problems discussed in this paper.}. Any of these composite fermions has a flavour index that goes from 1 to 3, as for the standard quarks. One  assumes that the Yukawa couplings of the heavy fermions as well as their mass terms\footnote{We omit the Yukawa terms $\tilde{\lambda}_U \bar{Q}^u_R H U_L + \tilde{\lambda}_D \bar{Q}^d_R \tilde{H} D_L + \text{h.c.}$, since they do not affect our considerations.}
\begin{equation}
\mathcal{L}_s = \lambda_U \bar{Q}^u_L H U_R + \lambda_D \bar{Q}^d_L \tilde{H} D_R
+ M_Q^u \bar{Q}^u_L Q^u_R + M_Q^d \bar{Q}^d_L Q^d_R +
M_U \bar{U}_L U_R + M_D \bar{D}_L D_R + \text{h.c.}
\label{Ls}
\end{equation}
respect a $U(3)_U\times U(3)_D$ flavour symmetry. Furthermore, to minimize possible new flavour effects and adopting the nomenclature of \cite{Redi:2011zi}, one assumes that this symmetry is extended to the standard right-handed quarks ({\it Right-compositeness}) or to the left-handed ones ({\it Left-compositeness}).
Explicitly
\begin{equation}
\mathcal{L}_\text{mix}^{R\text{-comp}} = m_U  \bar{U}_L u_R + m_D \bar{D}_L d_R 
+ \bar{q}_L \hat{m}_u Q^u_R + \bar{q}_L \hat{m}_d Q^d_R + \text{h.c.}
\label{mixing3R}
\end{equation}
or
\begin{equation}
\mathcal{L}_\text{mix}^{L\text{-comp}} = m_U \bar{q}_L Q^u_R + m_D \bar{q}_L  Q^d_R 
+ \bar{U}_L \hat{m}_u u_R + \bar{D}_L \hat{m}_d d_R + \text{h.c.}\,,
\label{mixing3L}
\end{equation}
where only the mass terms with a {\it hat} break the flavour symmetries. Accordingly, with $\hat{m}_{u,d}$ set to zero, the full flavour symmetry in the two cases is respectively (with a self explanatory notation):
\begin{itemize}
\item Right-compositeness: $U(3)_q\times U(3)_{U+u}\times U(3)_{D+d}$\,,
\item Left-compositeness: $U(3)_{q+ U +D}\times U(3)_{u}\times U(3)_{d}$\,.
\end{itemize}
In the two cases $\hat{m}_u$ and $\hat{m}_d$ transform under one of these symmetries as $(3,\bar{3},1)$ and $(3,1,\bar{3})$.

If one views the distinction between the third and the first two generations as a basic property of the entire theory, including the strong sector, which looks to us as a physically appealing point of view, it is straightforward to see how the $U(2)^3$ case works. Specifically, the analogue of $\mathcal{L}_s$ in (\ref{Ls}) only respects a $(U(2)\times U(1))_U\times (U(2)\times U(1))_D$ flavour symmetry, whereas the mixing Lagrangians in the two cases are:
\begin{equation}
\mathcal{L}_\text{mix}^{R\text{-comp}}(U(2)^3) \approx
m_U(A_u \, \CHuLbar \EHuR + B_u\, \CLuLbar\ELuR) +
\tilde m_U(a_u \,\EHqLbar \CHquR + b_u \,(\ELqLbar\V)\CHquR + c_u\, \ELqLbar\Delta Y_u \CLquR) +
\text{h.c.}
\label{mixing2R}
\end{equation}
or
\begin{equation}
\mathcal{L}_\text{mix}^{L\text{-comp}}(U(2)^3) \approx 
m_U(A_u\, \EHqLbar \CHquR + B_u \,\ELqLbar\CLquR) +
\tilde m_U( a_u\, \CHuLbar \EHuR + b_u\, (\CLuLbar\V)\EHuR + c_u \,\CLuLbar\Delta Y_u \ELuR) +
\text{h.c.}
\label{mixing2L}
\end{equation}
plus analogous terms for the down quarks. To clarify the notation, we always use capital letters for composite fields and, as in (\ref{BoldNotation}), vectors in the first two generations space are in boldface.
With $\tilde m_{U,D}$ switched off, the flavour symmetries are:
\begin{itemize}
\item Right-compositeness: $U(3)_q\times (U(2)\times U(1))_{U+ u}\times (U(2)\times U(1))_{D +d}$\,,
\item Left-compositeness: $(U(2)\times U(1))_{q+ U +D}\times U(3)_{u}\times U(3)_{d}$\,.
\end{itemize}
The terms only involving the fields of the third generation break these symmetries down to:
\begin{itemize}
\item Right-compositeness: $U(2)_q\times U(2)_{U+ u}\times U(2)_{D +d}$\,,
\item Left-compositeness: $U(2)_{q+ U +D}\times U(2)_{u}\times U(2)_{d}$\,,
\end{itemize}
up to overall baryon number, and, in analogy with  Section 2,
 $\V, \Delta Y_u, \Delta Y_d$ transform  under one of these symmetries as $(2,1,1),~(2,\bar{2},1),~(2,1,\bar{2})$ respectively.

\subsection{Flavour and CP violation}

The limited ability to calculate in a strongly interacting theory, like the one under discussion in this Section, makes it seem difficult to go beyond EFT considerations.  Due to the specific way of flavour breaking, by mass mixings as in (\ref {mixing3R}--\ref{mixing3L}) or in (\ref{mixing2R}--\ref{mixing2L}), this is actually not the case: only some of the flavour breaking operators survive among the ones discussed in Section~\ref{sec:eft}. As we now illustrate, this is summarized in Table~\ref{tab:u2u3} for  $U(3)^3$ and for  $U(2)^3$ respectively, with a neat distinction between Right- and Left-compositeness, and compared with generic EFT theory expectations.
We are interested in the relevant flavour violating operators involving the elementary fields to all orders in the strong interaction dynamics, i.e. after integrating out the heavy composite degrees of freedom.

\begin{table}[tb]
\renewcommand{\arraystretch}{1.5}
 \begin{center}
\begin{tabular}{lcccc}
\hline
& $b_L\leftrightarrow q_L$ && $s_L\leftrightarrow d_L$ & $b_R\leftrightarrow q_L$
\\\hline
$U(3)^3$ moderate $t_\beta$ & $\mathbbm R$ &$\leftrightarrow$& $\mathbbm R$ & $\mathbbm C$
\\
$U(3)^3$ $R$-comp. & $\mathbbm R$ &$\leftrightarrow$& $\mathbbm R$ & 0
\\
$U(3)^3$ $L$-comp. & 0 && 0 & 0
\\\hline
$U(2)^3$, $U(3)^3$ large $t_\beta$  & $\mathbbm C$ && $\mathbbm R$ & $\mathbbm C$
\\
$U(2)^3$ $R$-comp. & $\mathbbm C$ && $\mathbbm R$ & 0
\\
$U(2)^3$ $L$-comp. & $\mathbbm R$ && $\mathbbm R$ & $\mathbbm C$
\\\hline
\multirow{2}{100pt}{Relevant processes} &
$B_q^0$-$\bar B_q^0$ && $K^0$-$\bar K^0$ & $b\to s\gamma$
\\
& $b\to sl\bar l,\nu\bar\nu$ && $K\to\pi\nu\bar\nu$ & $b\to sl\bar l$
\\\hline
 \end{tabular}
 \end{center}
\caption{Expected new physics effects in $U(3)^3$ vs. $U(2)^3$ in the generic EFT cases and in the special cases of left- and right-compositeness, for chirality conserving and chirality breaking $\Delta F=1,2$ FCNC operators in the $B$ and $K$ systems. $\mathbbm R$ denotes possible effects, but aligned in phase with the SM, $\mathbbm C$ denotes possible effects with a new phase, and 0 means no or negligible effects. In $U(3)^3$, an additional feature is that the effects in $b\to q$ (where $q=d,s$) and $s\to d$ transitions are perfectly correlated.
In two Higgs doublet models at large $\tan\beta$, the possible effects in $U(3)^3$ correspond to the ones in $U(2)^3$ (4th row) and the correlation can be broken.
The last two rows show the processes most sensitive to the respective operators.
}
\label{tab:u2u3}
\end{table}

\subsubsection{Left-compositeness:  $U(3)^3$ case}

\begin{figure}[tb]
\centering
\includegraphics[width=\textwidth]{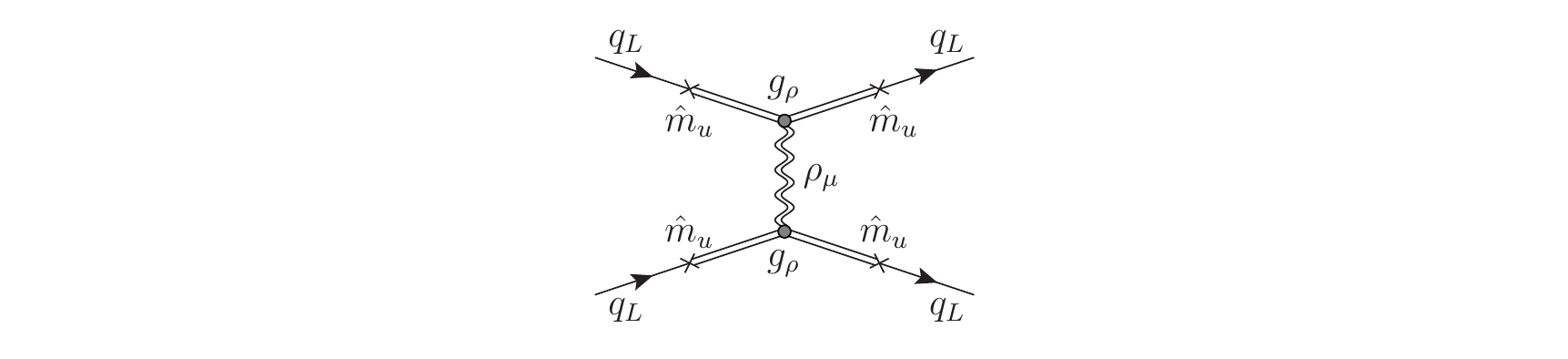}
\caption{Four-fermion operator contributing to the $\Delta F = 2$ amplitudes in the $U(3)^3$ case with Right-compositeness.\label{fig4}}
\end{figure}

 By pure symmetry considerations, several fermion bilinears can be written that would lead to flavour violation, e.g. $\bar{q}_L\hat{m}_u \hat{m}_u^\dagger  \gamma_\mu q_L$. However an operator containing this  bilinear cannot be generated without propagating the elementary field $u_R$ in an internal line, as seen from (\ref{mixing3L}).
On the contrary, one can generate $\bar{u}_R\hat{m}_u^\dagger  \hat{m}_u \gamma_\mu u_R$ or $\bar{d}_R\hat{m}_d^\dagger   \hat{m}_d\gamma_\mu d_R$, none of which gives rise to flavour violation after going to the physical mass basis for the quarks. In $U(3)^3$ and Left-compositeness the only operators that lead to flavour violation and can be generated by integrating out heavy degrees of freedom are of  charged-current type, e.g. $(\bar{u}_R\hat{m}_u^\dagger   \hat{m}_d\gamma_\mu d_R) (H^T\sigma_2 D_\mu H)$.
This operator would contribute to correct the SM rate for $B \rightarrow \tau \nu$ but only by a negligibly small amount, unless weighted by an anomalously large dimensionless coefficient.

\subsubsection{Right-compositeness: $U(3)^3$ case}

Unlike the case of Left-compositeness, here the bilinear $\bar{q}_L\hat{m}_u \hat{m}_u^\dagger  \gamma_\mu q_L$ can be generated, e.g. by mixing effects on the current $\bar{Q}^u\gamma_\mu Q^u$ coupled to a heavy $\rho$-like composite vector. Hence flavour changing operators of chirality conserving type do exist, as in the generic EFT case. 
From the diagram of Fig.~\ref{fig4} we estimate for the coefficient of the $\Delta F = 2$ operator
\begin{equation}
\mathcal{L}^{\Delta F = 2} = \frac{g_{\rho}^2}{M_{\rho}^2\lambda_U^4\sin^4\theta_R}\xi_{ij}^2(\bar q_{Li}\gamma_{\mu}q_{Lj})^2,
\end{equation}
where $\sin\theta_R = m_U / \sqrt{m_U^2 + M_U^2}$.
The only special property in composite Higgs models with $U(3)^3$ flavour symmetry is that no new phase appears in these operators, even for large $\tan{\beta}$, since bilinears like $\bar{q}_L\hat{m}_u \hat{m}_u^\dagger  \hat{m}_d \hat{m}_d^\dagger  \gamma_\mu q_L$ cannot be generated without propagation of the elementary fermions. Similarly no flavour violating bilinear of  chirality breaking type, like e.g. $\bar{q}_L\hat{m}_u \hat{m}_u^\dagger  \hat{m}_d d_R$, can be generated.

\subsubsection{Left-compositeness:  $U(2)^3$ case}

\begin{figure}[tb]
\centering
\subfigure[Right-handed compositeness \label{uno}]{
\includegraphics[width=\textwidth]{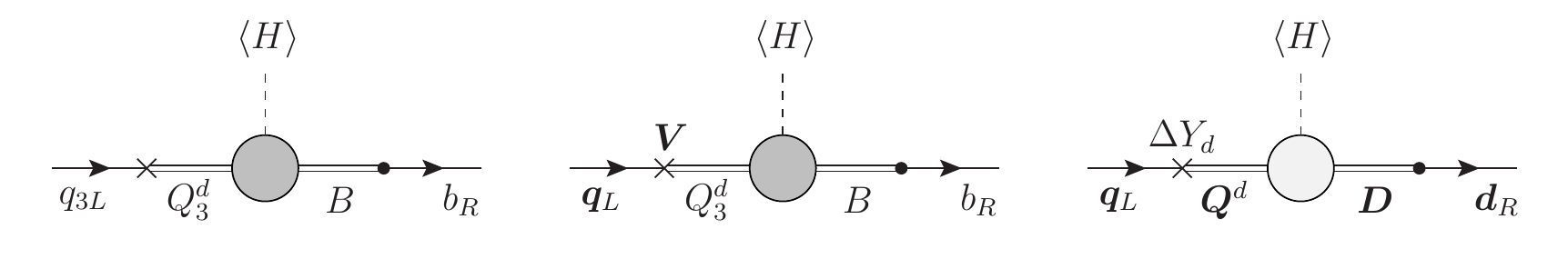}}
\subfigure[Left-handed compositeness \label{due}]{
\includegraphics[width=\textwidth]{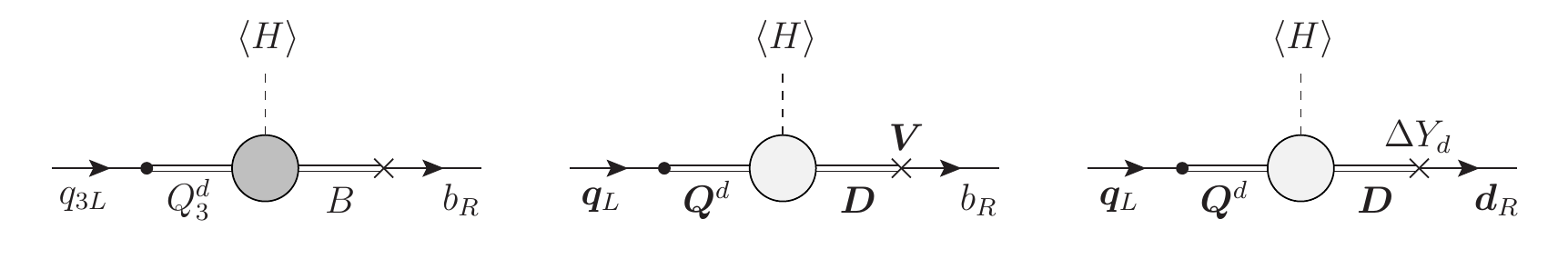}}
\caption{Leading contributions to chirality breaking bilinears (mass terms) from insertions of composite fermions, in the two cases of right- and left-handed compositeness. Crosses denote the mixings that break flavour symmetry, while dots denote the diagonal mixings. The strong dynamics in the first two diagrams in $(a)$ is the same, as it is in the last two diagrams in $(b)$. \label{A}}
\end{figure}

\begin{figure}[tb]
\centering
\includegraphics[width=\textwidth]{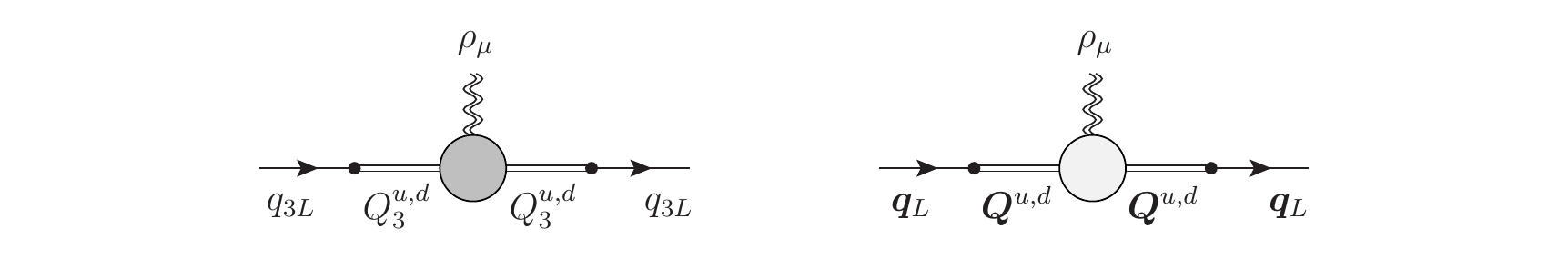}
\caption{Leading contribution to flavour conserving bilinears, in the case of left-handed compositeness. Flavour violation arises after rotating to the physical quark basis.\label{B}}
\end{figure}

The different treatment of the third generation quarks, elementary or composite, with respect to the first two has important consequences. Let us consider first the chirality breaking bilinears. In Fig.~\ref{A} we show the leading terms contributing to the effective Yukawa couplings of the down quarks to the Higgs boson. The important thing to note is that the strong Yukawa couplings of the Higgs doublet to the heavy quarks, although flavour conserving, are different if the coupling is to the third generation than if it is to the first two. Let us now consider the analogous diagrams with a photon line also attached to the strong Yukawa-coupling vertex, i.e. an operator contributing, when the Higgs boson gets a vev, to the dipole moments of the heavy quarks. Here too the dipole moment of the third-generation composite quarks is in general different from the one of the first two generations, but, what is more important, the two dipole moments will not be in the same ratio as the corresponding  Yukawa couplings. In turn this produces a misalignment in flavour space of the mass versus the magnetic moment operator, i.e. to a residual flavour breaking interaction with the size expected in the generic EFT approach and $\Lambda \approx 4\pi v\approx 3$ TeV, except that the presence of a new phase requires CP violation in the strong sector.

The situation is analogous in chirality conserving bilinears, where the diagrams in Fig.~\ref{B}  produce an effective current with a mismatch between the coefficient of the $\EHqLbar\gamma_\mu \EHqL$ term and the one of the $\ELqLbar\gamma_\mu \ELqL$ term. In turn, after going to the physical quark basis for the down quarks, this gives rise to flavour violation as in the generic EFT case, although without a new phase. 

\subsubsection{Right-compositeness:  $U(2)^3$ case}

In the chirality breaking bilinears the difference with the Left-compositeness case is in the heavy mediating quarks in Fig.~\ref{A}. As a consequence the relative coefficients between the bilinears $\EHqLbar(1, \sigma_{\mu\nu})\EHdR$ and 
$\ELqLbar(1, \sigma_{\mu\nu})\EHdR$ is the same for the mass terms as for the dipole terms. The relative coefficient changes  for $\ELqLbar(1, \sigma_{\mu\nu})\ELdR$. Nevertheless, when going to the physical mass basis, the dipole terms are also diagonalized. On the contrary, for chirality conserving bilinears, there is no special change relative to the general EFT analysis.

\section{Lepton Flavour Violation}\label{sec:leptons}

\subsection{General setup and current bounds in Effective Field Theory}

If one tries to extend the considerations developed so far to the lepton sector, one faces two problems. First, while the hierarchy of charged lepton masses is comparable to the mass hierarchies in the quark sector, the leptonic charged-current mixing matrix  does not exhibit the hierarchical pattern of the CKM matrix. Secondly, in the likely possibility that the observed neutrinos are of Majorana type and are light because of a small mixing to heavy right-handed neutrinos, the relevant parameters in the Yukawa couplings of the  lepton sector,  $H \bar{l}_L Y_e e_R$ and $\tilde{H} \bar{l}_L Y_\nu \nu_R$,  are augmented, relative to the ones in the quark sector, by the presence of the right-handed neutrino mass matrix, $\nu_R^T M \nu_R$, which is unknown. To overcome these problems we make the following two hypotheses: \begin{itemize} \item we suppose that the charged leptons, thorough  $Y_e$, behave in a similar way to the quarks, whereas $Y_\nu$ and $M$ are responsible for the anomalously large neutrino mixing angles; \item we assume that $Y_\nu$ has no significant influence on flavour physics near the Fermi scale in spite of the presence at this scale of new degrees of freedom carrying flavour indices, like sneutrinos or heavy composite leptons. One can imagine many reasons for this to be the case, like e.g. in the discussion of the next Section.\end{itemize}

Extending $U(3)^3$ and $U(2)^3$ to the leptonic sector, we consider respectively a $U(3)_l\times U(3)_e$ and a $U(2)_l\times U(2)_e$ symmetry. Here comes another significant difference between the two cases: in the $U(3)_l\times U(3)_e$ case, with $Y_e$ transforming as a $(3, \bar{3})$, there is no new flavor changing phenomenon at the Fermi scale other than the  leptonic charged-current mixing matrix, since $Y_e$  can be diagonalized by a $U(3)_l\times U(3)_e$  transformation.
On the contrary, let us assume that the  $U(2)_l\times U(2)_e$ symmetry be  broken {\em weakly} by the spurions
\begin{equation}
\Delta Y_e = (2,\bar 2) \,,
\qquad
\Ve = (2,1) \,.
\label{lept_spur}
\end{equation}
By proceeding in the same way as for the quarks in Section 2.1, one can set
\begin{equation}
\Ve^T = (0, \eta), ~~ \Delta Y_e = R_{12}^e \Delta Y_e^{diag}
\end{equation}
and see the occurrence of flavour changing bilinears with two important differences relative to the quark case. Firstly, one cannot relate the size of $\eta$ or of the angle $\theta_e$ in $R_{12}^e$ to the mixing matrix in the leptonic charged current. Secondly, due to the importance of $\mu \rightarrow e \gamma$, one has to include in the expansion of the 
chirality breaking bilinears, analogue of eq. (\ref{cb_d}), subleading terms like $(\ELlLbar \Ve) (\Ve^\dagger \Delta Y_e \ELeR)$, and perform the rotation to the physical basis to the corresponding order in the spurions.

The flavour changing dimension six effective operators in the lepton sector can then be written as
\begin{enumerate}[i)]
\item $\tau\to\mu,e$, chirality breaking:
\begin{equation}
c_\tau \zeta_{i\tau} m_\tau \left(\bar{e}_{Li}\sigma_{\mu \nu} \tau_R\right) e F_{\mu\nu},
\label{taumue}
\end{equation}
\item $\mu\to e$, chirality breaking:
\begin{equation}
c_\mu \zeta_{e\mu} m_\mu \left(\bar{e}_{L}\sigma_{\mu \nu} \mu_R\right) e F_{\mu\nu},
\label{mue}
\end{equation}
\item $\tau\to\mu,e$, chirality conserving:
\begin{equation}
c^\beta_\tau \zeta_{i\tau} \left(\bar{e}_{Li}\gamma_\mu \tau_L\right) O^\beta_\mu, ~~
\end{equation}
\begin{equation}
O^\beta_\mu = \left(\bar{l}_L\gamma_\mu l_L\right),~\left(\bar{e}_R\gamma_\mu e_R\right),
~\left(\bar{q}_L\gamma_\mu q_L\right),~\left(\bar{u}_R\gamma_\mu u_R\right),~\left(\bar{d}_R\gamma_\mu d_R\right),
~\left(H^\dagger D_\mu H\right),
\end{equation}
\item $\mu\to e$, chirality conserving:
\begin{equation}
c^\beta_\mu  \zeta_{e\mu}\left(\bar{e}_{L}\gamma_\mu \mu_L\right)O^\beta_\mu, ~~
\end{equation}
\end{enumerate}
where $\zeta_{ij}= U_{eL}^{3i*}U_{eL}^{3j}$ and $|U_{eL}|\simeq R_{12}^eR_{23}^e$ is the left-handed charged lepton Yukawa diagonalization matrix.
All these coefficients are  model dependent and, in principle, of similar order.

\begin{table}[tbp]
\begin{center}
\renewcommand{\arraystretch}{1.5}
\renewcommand\tabcolsep{5pt}
\begin{tabular}{ccccccccc}
\hline
$\mu\to e\gamma$ & $2.4\times10^{-12}$ &\cite{Adam:2011ch}& $\mu\to 3e$ & $1.0\times10^{-12}$ &\cite{Bellgardt:1987du}& $\mu\to e \text{ (Ti)}$ & $6.1\times10^{-13}$ &\cite{Wintz:1998rp} \\
$\tau\to e\gamma$ & $3.3\times10^{-8}$ &\cite{:2009tk}& $\tau\to 3e$ &$2.7\times10^{-8}$ &\cite{Hayasaka:2010np}&&\\
$\tau\to \mu\gamma$ & $4.3\times10^{-8}$ &\cite{:2009tk}& $\tau\to 3\mu$ & $2.1\times10^{-8}$ &\cite{Hayasaka:2010np}&&\\
\hline
\end{tabular}
\end{center}
\caption{90\% C.L. experimental upper bounds on the branching ratios of 6 LFV decays and on the $\mu\to e$ conversion rate in Titanium.}
\label{tab:lfvexp}
\end{table}

\begin{table}[tbp]
\begin{center}
\renewcommand{\arraystretch}{1.5}
\begin{tabular}{lcccl}
\hline
operator & $|\widetilde c_{e\mu}|$ & $|\widetilde c_ {e\tau}|$ & $|\widetilde c_ {\mu\tau}|$ & constrained from\\
\hline
$m_{e_j} (\bar{e}_{Li}\sigma_{\mu \nu} e_{Rj}) e F_{\mu\nu}$    & 0.07 & 0.79 & 0.2 & $l_j\to l_i\gamma$\\
$(\bar{e}_{Li} \gamma_\mu e_{Lj})(\bar e_{Li} \gamma^\mu e_{Li})$ & 0.6 & 9.4 & 1.8 & $l_j\to 3l_i$ \\
$(\bar{e}_{Li} \gamma_\mu e_{Lj})(\bar e_{Ri} \gamma^\mu e_{Ri})$ & 0.9 & 13 & 2.6 & $l_j\to 3l_i$ \\
$(\bar{e}_{Li} \gamma_\mu e_{Lj})(\bar u \gamma^\mu u)$  & 0.03 &--&-- & $\mu\to e \text{ (Ti)}$\\
$(\bar{e}_{Li} \gamma_\mu e_{Lj})(\bar d \gamma^\mu d)$  & 0.03 &--&-- & $\mu\to e \text{ (Ti)}$\\
\hline
\end{tabular}
\renewcommand{\arraystretch}{1.0}
\end{center}
\caption{90\% C.L. upper bounds on the reduced coefficients defined in (\ref{eq:ctilde}). The last column lists the processes giving the strongest constraint on the respective operators.}
\label{tab:lfvbounds}
\end{table}

The current bounds on LFV processes are collected in Tab.~\ref{tab:lfvexp}. Using them, bounds can be set on the above coefficients, making assumptions about the mass scale of new physics and the size of the mixing angles $\zeta_{ij}$. Details on the dependence of the observables on the coefficients can be found in Appendix~\ref{sec:applfv}. In Tab.~\ref{tab:lfvbounds}, we show the bounds on the dimensionless reduced coefficients
\begin{equation}
\widetilde c_{ij} = c_{j} \times \left[\frac{3\,\text{TeV}}{\Lambda}\right]^2\left[\frac{\zeta_{ij}}{V_{ti}V_{tj}^*}\right],
\label{eq:ctilde}
\end{equation}
where the indices $i,j = e, \mu, \tau$ refer to the specific flavour transitions.
For $\Lambda = 3$ TeV some of these bounds are significant, especially from $\mu$-decay processes. Note however that the normalization of  the $\zeta_{ij}$ to the corresponding products of the  CKM matrix elements should only be taken as indicative. Note furthermore that the operators contributing to $\mu\to e$ conversion are suppressed in specific models by explicit gauge coupling factors (supersymmetry) or by small mass mixing terms (composite Higgs models, see below).

A specific case that can be studied along these lines is supersymmetry.
In that case, only the dipole operator in Eq.~(\ref{taumue},~\ref{mue}) is relevant in the processes listed in Tab.~\ref{tab:lfvbounds}. The dominant contribution (even for moderate $\tan\beta$), assuming heavy first and second generation sleptons and a common mass $\widetilde m$ for the charginos and third generation sleptons, reads
\begin{equation}
c_j \zeta_{ij} = \frac{g^2}{128\pi^2}\frac{W_L^{i3*}W_L^{j3}}{\widetilde m^2}\tan\beta \,,
\end{equation}
where $W_L$ governs lepton flavour violation in the charged lepton-slepton-gaugino vertices, $[\bar l_L^iW_L^{ij}\tilde l_L^j]\tilde W$. One then finds
\begin{align}
\frac{|W_L^{23*}W_L^{13}|}{|V_{ts}V_{td}^*|}
&< 0.6\times
\left[\frac{m_{\tilde\tau_L}}{500\,\text{GeV}}\right]^2
\left[\frac{10}{\tan\beta}\right]
,\\
\frac{|W_L^{33*}W_L^{13}|}{|V_{tb}V_{td}^*|}
&< 1.2\times
\left[\frac{m_{\tilde\tau_L}}{500\,\text{GeV}}\right]^2
\left[\frac{10}{\tan\beta}\right]
,\\
\frac{|W_L^{33*}W_L^{23}|}{|V_{tb}V_{ts}^*|}
&< 0.3\times
\left[\frac{m_{\tilde\tau_L}}{500\,\text{GeV}}\right]^2
\left[\frac{10}{\tan\beta}\right]
.
\end{align}
More details on the SUSY case can be found in Appendix~\ref{sec:appsusylfv}.

\subsection{LFV in composite Higgs models and the $g - 2$ of the muon}

To make a composite Higgs model fully realistic one must extend the discussion of Section~\ref{sec:composite} to the lepton sector as well. A rather unique way in which this can be done closely mimics the case of the quarks.  In the strong sector composite vector-like leptons, $L$, $E$ and $N$, are assumed to exist with the same quantum numbers of the elementary $l_L, e_R, \nu_R$ as well as Yukawa couplings and mass terms in analogy with (\ref{Ls})\footnote{Although we do not think it to be phenomenologically necessary,  to maximize the quark-lepton symmetry one can consider two $SU(2)_L$-doublets, $L^e$ and $L^\nu$.}. Similarly there will be mass mixing terms between the composite and the elementary fermions. The only asymmetry with the quark sector is in the presence of a  mass matrix for the elementary $\nu_R$, with elements much larger than any other scale. 
As a consequence the neutrino spectrum, per generation, consists of one light neutrino, which can be arranged to have standard left-handed weak interaction to a sufficient level of accuracy, two quasi-Dirac  neutrinos at the typical compositeness scale and one superheavy almost pure right-handed neutrino. Following the discussion of the previous Section we assume that the right-handed neutrino mixing matrix, $\bar{N}_L \hat{m}_\nu \nu_R$, and the Majorana mass matrix of the right-handed neutrinos have a flavour structure such that the mass matrix of the light neutrinos  gives rise to the large mixing angles of the leptonic charge current.

What about the flavour properties of the light charged leptons to all orders in the strong interactions? Let us consider first the case in which the strong sector conserves  a diagonal leptonic $U(3)_L$ symmetry. As mixing terms we can consider:
\begin{equation}
\mathcal{L}^{R\text{-comp}}_\text{mix} = m_E \bar{E}_L e_R + \bar{l}_L \hat{m}_e L_R
\end{equation}
or
\begin{equation}
\mathcal{L}^{L\text{-comp}}_\text{mix} = m_E \bar{l}_L L_R + \bar{E}_L \hat{m}_e e_R,
\end{equation}
with $\hat{m}_e$ transforming as $(3, \bar{3})$ under $U(3)_l\times U(3)_{L+e}$ or $U(3)_{L+ l}\times U(3)_e$ respectively. 
As anticipated, in either case there is no leptonic flavour changing phenomenon at the Fermi scale other than the standard mixing in the leptonic charged-current interaction. This is because $\hat{m}_e$ can be set to diagonal form and $\hat{m}_\nu$ has no effect at the Fermi scale.
The discussion of lepton flavour violation in the case of a $U(2)$ symmetry proceeds along similar lines as in the quark case. The strong interaction Lagrangian respects a $(U(2)\times U(1))_L$ symmetry whereas the mixing Lagrangians are:
\begin{equation}
\mathcal{L}^{R\text{-comp}}_\text{mix}(U(2)) \approx  m_E(A_e \, \CHeLbar \EHeR + B_e\, \CLeLbar\ELeR) +
\tilde m_E(a_e \,\EHlLbar \CHlR + b_e \,(\ELlLbar\Ve)\CHlR + c_e\, \ELlLbar\Delta Y_e \CLlR) +
\text{h.c.}
\end{equation}
or
\begin{equation}
\mathcal{L}^{L\text{-comp}}_\text{mix}(U(2)) \approx m_E(A_e\, \EHlLbar \CHlR + B_e \,\ELlLbar\CLlR) +
\tilde m_E( a_e\, \CHeLbar \EHeR + b_e\, (\CLeLbar\Ve)\EHeR + c_e \,\CLeLbar\Delta Y_e \ELeR) +
\text{h.c.}
\end{equation}
Both in the case of Left- and of Right-compositeness there are flavour violating, chirality conserving transitions, whereas a difference exists for chirality breaking transitions, as in the quark case. This means that the leading order operators, $\tau \rightarrow \mu, e + \gamma$ as in (\ref{taumue}), are present only in the Left-compositeness case, with an amplitude proportional to $m_\tau$. On the contrary, as a subleading phenomenon in the expansion in $\Delta Y_e$ and  $
\Ve$, the  $\mu \rightarrow  e + \gamma$ operator as in (\ref{mue}) exists both in Left- and Right-compositeness.

As an aside remark we note that a magnetic moment operator of the composite charged leptons 
\begin{equation}
\dfrac{\lambda_L v}{\sqrt{2} M^2} (\bar{L}\sigma_{\mu\nu} E) e F_{\mu\nu}
\end{equation}
 gives rise, after mass mixing, to a magnetic moment operator for the standard charged leptons 
\begin{equation}
\frac{1}{M^2} (\bar{l}_i \sigma_{\mu\nu} m_i e_i) e F_{\mu\nu}, ~~~ i = e, \mu, 
\end{equation}
where $M$ is a typical compositeness scale and $m_i$ are the masses of the standard charged leptons. In turn this corrects the $g-2$ anomalies by an extra contribution:
\begin{equation}
\Delta a_i \equiv \dfrac{\Delta (g - 2)_i}{2} = \frac{4\, m_i^2}{M^2}.
\end{equation}
This could explain the putative discrepancy between theory and experiment of the muon anomaly, $\Delta a_\mu \approx 3 \cdot10^{-9}$\cite{Prades:2009qp}, with a mass $M\approx 4$ TeV, while being consistent with the current information on the electron $(g-2)$. The exact correlation between the electron and the muon anomalies is a consequence of a $U(3)_L$   or a $(U(2)\times U(1))_L$ symmetries of the strong interaction Lagrangian. 

\section{Summary and outlook}

The motivation for the occurrence of new physics in the TeV range related to EWSB  also calls for the existence of new flavour physics phenomena at similar energies. For this to be the case, however, as indeed generally implied by motivated extensions of the SM, some physical mechanism must be operative to avoid the presence of unseen large deviations from the CKM picture of flavour and CP violation: if present at all they must be kept under some control. That this physical mechanism be related to the approximate $U(2)^3$ flavour symmetry observed in the quark sector of the SM Lagrangian is a hypothesis well worth of investigation, as pursued in this paper building on  previous work  in the special context of supersymmetry. We have also crucially assumed  that the breaking of  $U(2)^3$ takes place along definite directions, specified by (\ref{directions}) and motivated by minimality.

The first  conclusion that we draw from general Effective Field Theory considerations is apparent from Fig.~\ref{fig:DF2fits} and \ref{fig:DF1fits}: the wealth of current flavour physics data is still broadly compatible with new phenomena hiding at an interesting scale of about 3~TeV,  a clearly relevant energy range in beyond-the-SM theories of EWSB, either perturbative or strongly interacting. We find this a particularly significant fact from many different points of view:  most of all it gives  reasons to think that new physics searches in the flavour sector may be about to explore an interesting realm of phenomena. Note that, taking EFT considerations at face value, standard MFV based on $U(3)^3$ implies stronger constraints than in the $U(2)^3$ case, related to the differences summarized in Table~\ref{tab:u2u3}.

The relation of these considerations with models of EWSB is manifest. Both in supersymmetry and in composite Higgs models it is non trivial how specific  realizations cope with flavour and CPV tests other than playing with parameters. 
$U(2)^3$ may provide a more fundamental underlying reason. In the case of supersymmetry this has been shown elsewhere \cite{Barbieri:2011ci,Barbieri:2011fc}, with possible interesting connections with the squark spectrum under study at LHC. 
Here we have seen how  the $U(2)^3$ symmetry and its breaking can be implemented in a generic composite Higgs model and we have analyzed its peculiar consequences  for flavour and CPV, summarized in Table~\ref{tab:u2u3}. We defer to later work a detailed analysis of the phenomenological consequences for flavour violating as well as flavour conserving observables, as done e.g. in \cite{Buras:2011ph} in the general case and in \cite{Redi:2011zi,Domenech:2012ai} for the $U(3)^3$ case.

Lepton Flavour Violation is in many respects the next due subject although with a major difficulty: the peculiar properties of the neutrino mixing matrix, quite different from the quark one, and, perhaps not unrelated, the weaker information available in the lepton sector relative to the quark sector due to the possible role of the right-handed neutrino mass matrix. Altogether we find it conceivable that the observed neutrino masses and mixings have a very high energy origin with little impact on Fermi scale physics. Taking this view, we have on the contrary assumed that  the charged leptons may behave with respect to flavour in a similar way to the quarks, with a natural extension of  $U(2)^3$ to a $U(2)_l\times U(2)_e$. 
While this can only be a qualitative picture, since we lack any direct information on the relevant mixing matrix, it gives nevertheless a possible coherent description of LFV signal in the TeV range. One interesting feature characteristic of $U(2)_l\times U(2)_e$, suitably broken as in (\ref{lept_spur}), is in the comparison between the $\mu\rightarrow e$ and the $\tau\rightarrow \mu$ transitions, with the chirality breaking operators respectively proportional to $\zeta_{\mu \tau} m_\tau$ and $\zeta_{e\mu} m_\mu$.
As for $U(2)^3$ in the quark case, the implementation of the  $U(2)_l\times U(2)_e$ in a generic composite Higgs model is possible with peculiarities that  distinguish Left- from Right-compositeness. Even in this case the associated phenomenology deserves further study inside and outside the flavour sector.

\section*{Acknowledgments}
This work was supported by the EU ITN ``Unification in the LHC Era'', 
contract PITN-GA-2009-237920 (UNILHC) and by MIUR under contract 2008XM9HLM.

\appendix

\section{CKM matrix and effective operators}\label{sec:yukawa}

To a sufficient approximation, the general chirality conserving quark bilinear of eq. \eqref{cc} gives rise to the kinetic term
\begin{equation}
\bar q_L\slashed{\partial}X_{\rm kin}q_L = a\,\EHqLbar\slashed{\partial}\EHqL + b\,\ELqLbar\slashed{\partial}\ELqL + c\,(\ELqLbar\V)\slashed{\partial}\EHqL + c^*\,\EHqLbar\slashed{\partial}(\V^{\dag}\ELqL) + d(\ELqLbar\V)\slashed{\partial}(\V^{\dag}\ELqL),
\end{equation}
as well as to all the possible chirality conserving interaction terms
\begin{equation}
\bar q_L\gamma^{\mu}X^{\alpha}_{\rm int}q_L = a^{\alpha}\EHqLbar\gamma^{\mu}\EHqL + b^{\alpha}\ELqLbar\gamma^{\mu}\ELqL + c^{\alpha}(\ELqLbar\V)\gamma^{\mu}\EHqL + c^{\alpha*}\EHqLbar\gamma^{\mu}(\V^{\dag}\ELqL) + d^{\alpha}(\ELqLbar\V)\gamma^{\mu}(\V^{\dag}\ELqL),
\end{equation}
where $q_L = (\ELqL, \EHqL)$, and $\alpha$ labels the different interactions. All the parameters except the $c$'s are real by hermiticity. Chirality conserving interactions among the right-handed quarks are also possible, but there flavour violations are at higher orders in the spurions (e.g. $\EHdRbar\gamma^{\mu}\V^{\dag}\Delta Y_d\ELdR$) and thus negligible.

The chirality breaking bilinears of eq.s \eqref{cb_d} and \eqref{cb_u} generate the Yukawa interactions, and thus the mass term for the up- and down-type quarks
\begin{align}
\bar q_L Y_u u_R &= \lambda_t ( \EHqLbar\EHuR + x_t\ELqLbar\V\EHuR + \ELqLbar\Delta Y_u \ELuR), &(u\leftrightarrow d, t\leftrightarrow b),
\end{align}
where we have reabsorbed the $a$'s and $c$'s of eq.s \eqref{cb_d}, \eqref{cb_u} into the $\lambda$'s and the spurions to match the definitions of \cite{Barbieri:2011ci}; they generate also the $\sigma_{\mu\nu}$-terms
\begin{align}
\bar q_L \sigma_{\mu\nu}\mu_u^{\beta} u_R &= \lambda_t (a_u^{\beta} \EHqLbar\sigma_{\mu\nu}\EHuR + b_u^{\beta}\ELqLbar\V\sigma_{\mu\nu}\EHuR + c_u^{\beta}\ELqLbar\Delta Y_u \sigma_{\mu\nu}\ELuR), &(u\leftrightarrow d, t\leftrightarrow b).
\end{align}
As already stated in Section \ref{sec:eft}, after suitable $U(2)^3$ transformations the spurions take the form of eq. \eqref{spurions}. Moreover, $\lambda_t$, $\lambda_b$, $c$ and either $x_t$ or $x_b$ can be made real through phase redefinitions of $\EHuR$, $\EHdR$, $\EHqL$ and $\ELqL$, respectively. In this basis the previous operators can be written in the more compact form
\begin{align}\label{ccmatrix}
X_{\rm kin} &= A\mathbbm{1} + BR_{23}\I_{32} R_{23}^{T},\\
X_{\rm int}^{\alpha} &= A^{\alpha}\mathbbm{1} + B^{\alpha}U_{23}^{\alpha}\I_{32} (U_{23}^{\alpha})^{\dag},
\end{align}
where $\I_{32} = {\rm diag}(0, O(\epsilon^2), 1)$ and the $A$'s, $B$'s and $C$'s are real parameters, and
\begin{align}
Y_u &= \lambda_t (R_{23}^u\I_3 + R_{12}^u\Delta \tilde Y^{\rm diag}_u), & Y_d &= \lambda_b (U_{23}^d\I_3 + U_{12}^d\Delta \tilde Y^{\rm diag}_d),\\
\mu_{u}^{\beta} &= \lambda_t(a_u^{\beta} V_{23}^u\I_3 + c_u^{\beta} R_{12}^u\Delta \tilde Y^{\rm diag}_u), & \mu_{d}^{\beta} &= \lambda_b(a_d^{\beta} V_{23}^d\I_3 + c_d^{\beta} U_{12}^d\Delta \tilde Y^{\rm diag}_d),
\end{align}
where $\I_3 = {\rm diag}(0,0,1)$, $\Delta \tilde Y_{u,d}^{\rm diag} = {\rm diag}(\epsilon_1^{u,d}, \epsilon_2^{u,d},0)$ are defined as in \eqref{cbdown} and we have implicitly chosen $x_t$ to be real. Here and in the following $U_{ij}$ and $V_{ij}$ stand always for unitary matrices in the $(i,j)$ sector, while $R_{ij}$ indicate orthogonal matrices.

We want to derive the expressions for these operators in the physical basis where the quark masses are diagonal, and the kinetic term has the canonical form.
The kinetic term is diagonalized by a real rotation in the $(2,3)$ sector, and it takes the canonical form after wavefunction renormalizations of the fields. One can check that  these transformations do not alter, to a sufficient accuracy, the structure of the other operators, but cause only $O(1)$ redefinitions of the parameters.

The mass terms are diagonalized approximately by the transformation
\begin{equation}
Y_u\mapsto (R_{12}^u)^T(R_{23}^u)^T Y_u\equiv (R_L^u)^T Y_u,\qquad Y_d\mapsto (U_{12}^d)^{\dag}(U_{23}^d)^{\dag} Y_d\equiv (U_L^d)^{\dag} Y_d,
\end{equation}
up to right-handed transformations of order $\epsilon\epsilon_{1,2}^{u,d}$. Therefore one goes to the physical basis for the quarks by
\begin{align}
u_L&\mapsto R_L^u u_L, & d_L&\mapsto U_L^d d_L,
\end{align}
and the Cabibbo-Kobayashi-Maskawa matrix, defined in \eqref{CKM}, is
\begin{equation}
V \simeq (R_{12}^u)^T(R_{23}^u)^TU_{23}^d U_{12}^d\equiv (R_{12}^u)^T U_{23}^{\epsilon} U_{12}^d,
\end{equation}
where $U_{23}^{\epsilon}$ is a unitary transformation of order $\epsilon$. If the parameters $x_t$, $x_b$ are both real (e.g. if there are no $CP$ phases outside the spurions) then $U_{23}^d$ is real, and also $U_{23}^{\epsilon}$ becomes a real rotation.

Finally, in the physical basis the chirality conserving operators become
\begin{equation}
X_{\rm int}^{\alpha}\mapsto A^{\alpha}\mathbbm{1} + B^{\alpha}(U_{12}^d)^{\dag} U_{23}^{\alpha}\I_{32}(U_{23}^{\alpha})^{\dag}U_{12}^d,
\end{equation}
and the $\sigma_{\mu\nu}$-terms
\begin{equation}
\mu_d^{\beta}\mapsto \lambda_b\big(a_d^{\beta}(U_{12}^d)^{\dag} U_{23}^{\beta}\I_3 + c_d^{\beta} \Delta\tilde Y_d^{\rm diag}\big),
\end{equation}
and similarly for the up sector.

\section{Up quark sector within $U(2)^3$}\label{sec:up}

In this appendix we study the contributions to flavour
and CP violation in the up quark sector arising in the $U(2)^3$ framework.
In particular, we concentrate on the contributions to $D$-$\bar{D}$ mixing, to direct CP violation in $D$-decays,
to the top decays $t\rightarrow c \gamma$ and $t\rightarrow c Z$ and to the top chromo-electric dipole moment
(CEDM).

Within our setup, the relevant effective operators for the above processes are:
\begin{align}
\mathcal H_{\rm LL}^{\rm D}  & = \frac{c_{LL}^D }{\Lambda^2}\, \xi_{uc}^2 \, \dfrac{1}{2}\left({\bar u}_L
\gamma_\mu  c_L \right)^2, \label{eq:Dmixing} \\
\mathcal H_{\rm cb}^{\rm D}  & = \frac{c_{g}^{D} e^{i \phi_{g}^{D}}}{\Lambda^2} \,m_c \,\xi_{uc}
\left( {\bar u}_L \sigma_{\mu \nu} c_R \right) g_s G_{\mu \nu},
\label{eq:Ddecay}\\
\mathcal H_{\rm cb}^{\rm t,\, \alpha} & = \frac{c_{\alpha}^{t} e^{i \phi_{\alpha}^{t}}}{\Lambda^2}
\,m_t \,\xi_{ct} \left( {\bar c}_L \sigma_{\mu \nu} t_R \right) O^\alpha_{\mu \nu},
\quad O^\alpha_{\mu \nu} = e F_{\mu \nu}, ~\frac{g}{c_w} Z_{\mu \nu},
\label{eq:topcb} \\
\mathcal H_{\rm cc}^{\rm t} & = \frac{c_\text{cc}^{t} e^{i \phi_\text{cc}^{t}}}{\Lambda^2}\frac{v^2}{2} \,\xi_{ct}
\left({\bar c}_L \gamma_\mu t_L \right) \frac{g}{c_w} Z_\mu,
\label{eq:topcc}\\
\mathcal H_{\rm dm}^t & = \frac{c_\text{dm} e^{i \phi_\text{dm}}}{\Lambda^2}\,m_t
\left( {\bar t}_L \sigma_{\mu \nu} t_R \right) g_s G_{\mu \nu},
\end{align}
where
$c_w = \cos \theta_w$ and $c_{LL}^{D}$, $c_g^D$, $c_{\alpha}^{t}$, $c_\text{cc}^{t}$,
$c_\text{dm}$ are real parameters, with the phases made explicit wherever present.
All these coefficients are model dependent and, in principle, can be of $O(1)$.
Since $\Lambda \gg v$, the requirement of $SU(2)_L$ invariance correlates $c_{LL}^{D}$, $c_\text{cc}^{t}$ and
$\phi_\text{cc}^{t}$ with the analogous parameters in the down sector. One can easily see they have to be equal
within a few percent, and so they have to respect bounds similar to those for $c_{LL}^K$, $c_H$ and
$\phi_H$ (see Figures \ref{fig:DF2fits} and \ref{fig:DF1fits}).

\subsection{$D$ mixing and decay}

In the neutral $D$ meson system the SM short distance contribution to the mixing is orders of magnitudes
below the long distance one, thus complicating the theoretical calculation of the mass and width splittings
$x$ and $y$ (see \cite{Gedalia:2009kh}, also for a discussion of the relevant parameters).
Despite the above uncertainties, many studies (see \cite{Falk:2001hx, Falk:2004wg} and references therein)
indicate that the standard model could naturally account for
the values $x \sim y \sim 1 \%$, thus explaining the measured $95\%$ CL intervals
$x \in [0.19,0.97]\%$, $y \in [0.54,1.05]\%$ \cite{Asner:2010qj}.

Here, like in \cite{Gedalia:2009kh, Isidori:2011qw}, we take the conservative approach
of using the above data as upper bounds to constrain new physics contributions.
Referring to the analysis carried out in \cite{Gedalia:2009kh, Isidori:2011qw}, within our framework it turns
out that the most effective bound is the one on the coefficient in the operator $\mathcal H_{\rm LL}^{\rm D}$.
In our notation it reads
\beq
{c_{LL}^D}^2 \left(\dfrac{3 \,\text{TeV}}{\Lambda}\right)^2 < 90,
\label{eq:Dmix}
\eeq
so to saturate it we would need values of $c_{LL}^D$ that are excluded, since they would imply a too large contribution to
$\Delta F = 2$ observables in the down sector.

Suppose now that the recent LHCb evidence for CP violation in $D$ decays \cite{Aaij:2011in}
is not due to the SM. The quantity of interest is the difference between the time-intagrated CP asymmetries in the decays
$D^0 \to K^+ K^-$ and $D^0 \to \pi^+ \pi^-$, for which the HFAG reports the world average $\Delta a_\text{CP} = a_{KK} - a_{\pi \pi} = -(0.645 \pm 0.180)$ \cite{Asner:2010qj}.
In reference \cite{Isidori:2011qw} all the possible effective
operators contributing to the asymmetry are considered, while respecting at the same time the bounds coming from $D$-$\bar{D}$ mixing and
from $\epsilon_K'/\epsilon_K$.
Following that analysis, the only operator that can give a relevant contribution in our setup is
$\mathcal H_{\rm cb}^{\rm D}$, $\Delta a_\text{CP}$ being proportional to the imaginary part of the relative coefficient.
Referring to the estimations carried out in \cite{Isidori:2011qw} and \cite{Giudice:2012qq} for the hadronic
matrix elements, to reproduce the measured value of $\Delta a_\text{CP}$ one would need
\beq
c_g^D \sin ({\rm arg} \xi_{uc} + \phi_g^D) \left( \frac{3\, \text{TeV}}{\Lambda}\right)^2 \simeq 40,
\label{eq:Ddevay}
\eeq
a value out of reach if we want to keep the parameter $c_g^D$ to be of order one.

\subsection{Top FCNC and dipole moments}

The LHC sensitivity at 14 TeV with $100 \, \text{fb}^{-1}$ of data is expected to be (at $95 \%$ CL)
\cite{Carvalho:2007yi}:
$\text{BR}(t \to c\, Z, \,u\, Z) \simeq 5.5 \times 10^{-5}$ and $\text{BR}(t \to c\, \gamma,\, u\,\gamma)
\simeq 1.2 \times 10^{-5}$. Here we concentrate on the charm channels, since both in the SM and in our framework
the up ones are CKM suppressed. In the SM, $\text{BR}(t \to c\, Z,\, c\, \gamma)$ can be estimated to be
of order $ (m_b^2/m_W^2)^2 \, |V_{cb}|^2 \, \alpha^2/s_w^2 \sim 10^{-12}$, so that an experimental observation
will be a clear signal of new physics. To estimate the $U(2)^3$ effects for these processes, we
follow the analysis carried out in \cite{Fox:2007in}. The dominant contributions are those given by the operators
$\mathcal H_{\rm cb}^{\rm t,\, \gamma}$ for $t \to c \gamma$, and $\mathcal H_{\rm cb}^{\rm t,\, Z}$ and
$\mathcal H_{\rm cc}^{\rm t}$ for $t \to c Z$. We obtain
\begin{align}
\text{BR}(t \to c\, \gamma) \simeq & \; 1.7 \times 10^{-8} \left(\dfrac{3 \, \text{TeV}}{\Lambda}\right)^4
{c_{\gamma}^t}^2,\\
\text{BR}(t \to c\, Z) \simeq & \; 8.5 \times 10^{-8} \left(\dfrac{3 \, \text{TeV}}{\Lambda}\right)^4 
\left(0.61 \, {c_Z^t}^2 + 0.39 \, {c_\text{cc}^t}^2 + 0.83 \, c_Z^t \, c_\text{cc}^t \cos (\phi_\text{cc}^t - \phi_Z^t)\right),
\label{eq:topBR}
\end{align}
leading us to conclude that any non-zero evidence for these decays at the LHC could not be explained in our setup,
unless we allow the dimensionless coefficients to take values more than one order of magnitude bigger
than the corresponding ones in the down sector (actually this could be possible only for $c_Z^t$ and $c_{\gamma}^t$
but not for $c_\text{cc}^t$, because of its correlation with $c_H$ and of the bounds of Fig.~\ref{fig:DF1fits}).

The recent analysis carried out in \cite{Kamenik:2011dk} has improved previous bounds \cite{CorderoCid:2007uc}
on the top CEDM $\tilde{d}_t$
by two orders of magnitude, via previously unnoticed contributions of $\tilde{d}_t$ to the neutron
electric dipole moment (EDM).
In deriving this bound, the authors of \cite{Kamenik:2011dk} have assumed
the up and down quark EDMs $d_{u,d}$ and CEDMs $\tilde{d}_{u,d}$ to be negligible.
This is relevant in our context if
\begin{itemize}
 \item we allow for generic phases outside the spurions $\V$,
 $\Delta Y_u$ and $\Delta Y_d$,
 \item we assume that some other mechanism is responsible for making $d_{u,d}$ and $\tilde{d}_{u,d}$
 negligible. Notice that this is actually the case in SUSY with heavier first two generations, where
 on the contrary there is no further suppression of $\tilde{d}_t$ with respect to the EFT natural estimate.
\end{itemize}
Then,
the bound given in \cite{Kamenik:2011dk} imposes
\beq
c_\text{dm} |\sin \phi_\text{dm}|\, \left( \dfrac{3\, \text{TeV}}{\Lambda} \right)^2 < 0.6 \,,
\label{topCEDM}
\eeq
so that future experimental improvements in the determination of the neutron EDM
will start to challenge the $U(2)^3$ scenario with CP violating phases outside the spurions, if the
hypothesis of negligible $d_{u,d}$ and $\tilde{d}_{u,d}$ is realized.

\section{Lepton Flavour Violation in $U(2)^2$}\label{sec:applfv}

Here we list the contributions to LFV observables induced by the operators
\begin{align}
\mathcal H_\text{eff} =
&\sum_{j>i}\frac{\zeta_{ij}}{\Lambda^2}\bigg[
c_j m_{e_j} (\bar{e}_{Li}\sigma_{\mu \nu} e_{Rj}) e F_{\mu\nu}+
c_j^{l}(\bar{e}_{Li} \gamma_\mu e_{Lj})(\bar e_{Li} \gamma^\mu e_{Li})+
c_j^{e}(\bar{e}_{Li} \gamma_\mu e_{Lj})(\bar e_{Ri} \gamma^\mu e_{Ri})
\\
&+
c_j^{u}(\bar{e}_{Li} \gamma_\mu e_{Lj})(\bar u \gamma^\mu u)+
c_j^{d}(\bar{e}_{Li} \gamma_\mu e_{Lj})(\bar d \gamma^\mu d)
\bigg]
+\text{h.c.}
\end{align}
The $l^j\to l^i\gamma$ branching ratio is given by
\begin{equation}
 \text{BR}( l^j \rightarrow  l^i \gamma)
= \frac{192\pi^3 \alpha}{G_F^2} \frac{|\zeta_{ij} c_j|^2}{\Lambda^4}
\;b^{ij}\,,
\end{equation}
where $b^{ij}=\text{BR}( l^j\rightarrow l^i\nu\bar\nu)$.
The $ l^j \rightarrow  l^i\bar l^i l^i$ branching ratio reads \cite{Hisano:1995cp,Arganda:2005ji}
\begin{multline}
\text{BR}( l^j \rightarrow 3 l^i)=\frac{1}{2G_F^2}b^{ij}\frac{|\zeta_{ij}|^2}{\Lambda^4}
\bigg[
e^4|c_j|^2\left(16\ln\frac{m_{ l_j}}{m_{ l_i}}-22\right)
+\frac{1}{2}|c^l_j|^2
+\frac{1}{4}|c^e_j|^2
\\
+e^2\left(2c_j c^{l*}_j+c_j c^{e*}_j+\text{h.c.}\right)
\bigg] .
\end{multline}
For $\mu$-$e$ conversion, one obtains \cite{Barbieri:1995tw,Hisano:1995cp}
\begin{equation}
\Gamma(\mu\to e)=
\frac{\alpha^3}{4\pi^2}
\frac{Z_\text{eff}^4}{Z} |F(q)|^2 m_\mu^5
\frac{|\zeta_{ij}|^2}{\Lambda^4}
\left|
(2Z+N)c_j^u+(2N+Z)c_j^d
-
2Ze^2 c_j
\right|^2 \,.
\end{equation}
In the case of ${}^{48}_{22}$Ti, one has $Z_\text{eff}=17.6$ and $|F(q^2)|=0.54$ and the conversion rate is defined as 
\begin{equation}
\text{CR}(\mu\text{ Ti}\to e\text{ Ti}) =
\frac{\Gamma(\mu\text{ Ti}\to e\text{ Ti})}{\Gamma(\mu\text{ Ti}\to \text{capture})},
\end{equation}
where the capture rate is $\Gamma(\mu\text{ Ti}\to \text{capture})=(2.590 \pm 0.012)\times10^6 ~\text{s}^{-1}$.

\subsection{Radiative lepton decay in SUSY $U(2)^2$}\label{sec:appsusylfv}

In supersymmetry with a minimally broken $U(2)^2$ symmetry in the lepton sector, assuming first and second generation sleptons to be decoupled, the dominant contributions to the LFV dipole operators come from chargino-sneutrino and neutralino-stau loops. Neglecting trilinear terms, these can be written as
\begin{align}
\frac{64\pi^2}{\Lambda^2} \zeta_{ij}
c_j^{\rm chargino} &=  \frac{W_L^{i3*}W_L^{j3}}{m_{\tilde\tau_L}^2}\left[
 g^2|Z_+^{1a}|^2  \,F_c(x_{\tilde\chi^-_a}) - \sqrt{2}g\frac{m_{\tilde\chi^-_a}}{v}Z_-^{2a*}Z_+^{1a*}
\,G_c(x_{\tilde\chi^-_a})
\right],\\
\frac{64\pi^2}{\Lambda^2} \zeta_{ij}
c_j^{\rm neutralino} &=  \frac{W_L^{i3*}W_L^{j3}}{m_{\tilde\tau_L}^2}\bigg[
\frac{g^2}{2c_w^2}|Z_N^{1a}s_w+Z_N^{2a}c_w|^2  \,F_n(x_{\tilde\chi^0_a}) \\
&\qquad\qquad\qquad\qquad - \frac{g}{c_w}\frac{m_{\tilde\chi^0_a}}{v} Z_N^{3a*}(Z_N^{1a*}s_w+Z_N^{2a*}c_w)
\,G_n(x_{\tilde\chi^0_a})
\bigg],
\end{align}
where $x_{\tilde\chi}=m_{\tilde\chi}^2/m_{\tilde\tau_L,\tilde\nu_\tau}^2$. The matrices $Z_+$, $Z_-$ and $Z_N$ are defined as in \cite{Rosiek:1995kg}. The loop functions are
\begin{align}
F_c(x) &= -\frac{1}{12 (1-x)^3} \left( 2+5x-x^2 + 6x\frac{\ln x}{1-x} \right),
~F_c(1)=-\frac{1}{24}\,,\\
G_c(x) &= -\frac{1}{2 (1-x)^2} \left( -3+x - 2\frac{\ln x}{1-x} \right), 
~G_c(1)=-\frac{1}{3}\,,\\
F_n(x) &= \frac{1}{12 (1-x)^3} \left( 1-5x-2x^2 - 6x^2\frac{\ln x}{1-x} \right), 
~F_n(1)=\frac{1}{24}\,,\\
G_n(x) &= \frac{1}{2 (1-x)^2} \left( 1+x + 2x \frac{\ln x}{1-x} \right),
~G_n(1)=\frac{1}{6}\,.
\end{align}

\bibliographystyle{My}
\bibliography{u2general}

\end{document}